\documentclass{aa}  

\usepackage{natbib}
\usepackage{graphicx}
\usepackage{txfonts}
\usepackage{xcolor,colortbl}

\definecolor{Gray}{gray}{0.85}
\definecolor{LightCyan}{rgb}{0.88,1,1}
\newcolumntype{a}{>{\columncolor{Gray}}c}
\newcolumntype{b}{>{\columncolor{white}}c}
\usepackage[colorlinks=true,citecolor=blue]{hyperref}
\usepackage{float}
\usepackage{soul}

\makeatletter
\renewcommand*\aa@pageof{, page \thepage{} of \pageref*{LastPage}}
\makeatother

\begin{document} 
    \title{Chemical evolution of ultra-faint dwarf galaxies in the self-consistently calculated IGIMF theory}
    \titlerunning{Chemical evolution of UFD Bo\"otes~I in the IGIMF theory}

  \author{Zhiqiang Yan\inst{1} \fnmsep \inst{2}
          \and
          Tereza Jerabkova \inst{1} \fnmsep \inst{2} \fnmsep \inst{3} \fnmsep \inst{4} \fnmsep \inst{5}
          \and
          Pavel Kroupa \inst{1} \fnmsep \inst{2} \fnmsep 
          }

  \institute{Helmholtz-Institut f{\"u}r Strahlen- und Kernphysik (HISKP), Universität Bonn, Nussallee 14–16, 53115 Bonn, Germany
             \\ Emails: yan@astro.uni-bonn.de; tereza.jerabkova@eso.org; pkroupa@uni-bonn.de
         \and
             Charles University in Prague, Faculty of Mathematics and Physics, Astronomical Institute, V Hole{\v s}ovi{\v c}k{\'a}ch 2, CZ-180 00 Praha 8, Czech Republic
        \and 
            Astronomical Institute, Czech Academy of Sciences, Fri\v{c}ova 298, 25165, Ond\v{r}ejov, Czech Republic
        \and 
            Instituto de Astrof{\'i}sica de Canarias, E-38205 La Laguna, Tenerife, Spain
        \and
            GRANTECAN, Cuesta de San Jose s/n, 38712 Brena Baja, La Palma, Spain
             }

  \date{Received 24 January 2020 / Accepted 23 March 2020}

  \abstract
  {
    The galaxy-wide stellar initial mass function (gwIMF) of a galaxy in dependence of its metallicity and star formation rate (SFR) can be calculated by the integrated galactic IMF (IGIMF) theory. Lacchin et al. (2019) apply the IGIMF theory for the first time to study the chemical evolution of the ultra-faint dwarf (UFD) satellite galaxies and failed to reproduce the data. Here, we find that the IGIMF theory is naturally consistent with the data. We apply the time-evolving gwIMF calculated at each timestep. The number of type Ia supernova explosions per unit stellar mass formed is renormalised according to the gwIMF. The chemical evolution of Bo\"otes~I, one of the best observed UFD, is calculated. Our calculation suggests a mildly bottom-light and top-light gwIMF for Bo\"otes~I, and that this UFD has the same gas-consumption timescale as other dwarfs but was quenched about 0.1 Gyr after formation, being consistent with independent estimations and similar to Dragonfly 44. 
    The recovered best fitting input parameters in this work are not covered in the work of Lacchin et al. (2019), creating the discrepancy between our conclusions. In addition, a detailed discussion of uncertainties is presented addressing how the results of chemical evolution models depend on applied assumptions. This study demonstrates the power of the IGIMF theory in understanding the star-formation in extreme environments and shows that UDFs are a promising pathway to constrain the variation of the low-mass stellar IMF.
    }


\maketitle


\section{Introduction}\label{sec: Introduction}

Ultra-faint dwarf galaxies (UFDs) represent a continuous extension of dwarf galaxies in stellar mass, surface brightness, size, dynamical mass, and metallicity \citep{2019ARA&A..57..375S}. 
These fainter objects started to be observable with SDSS (and other modern surveys), the name UFD being first used by \cite{Willman2005} who speculated if the observed object is a dwarf galaxy or unusual globular cluster (GCs).
With more data it became evident that they are indeed dwarf galaxies (cf. \citealt{2011PASA...28...77F}) with a self-enriched stellar population formed over an extended period of $\gtrapprox 100$ Myr \citep{2013ApJ...767..134V,2014A&A...562A.146I,2015ApJ...799L..21W}.

The extreme physical properties of UFDs have led to many studies \citep{2013ApJ...763...61G,2014MNRAS.441.2815V,2014ApJ...796...11W,2015ApJ...807..154B,2015MNRAS.446.4220R,2016ApJ...826..110F,2017ApJ...848...85J} and discussions on their origin \citep[e.g.][]{2012LRR....15...10F,2018Natur.561E...4K}.
Dwarfs and UFDs with GC-like stellar masses and possibly a quick shut-off of further gas accretion with stars formed solely from the gas they had initially, point to a promising testing bed of chemical evolution models and the environmentally dependent variable galaxy-wide stellar initial mass function (gwIMF), as pioneered by \citet{2014MNRAS.437..994R}, \citet{2015MNRAS.449.1327V}, and \citet[version 1]{2019arXiv191108450L}. 

Most recently, \cite{2019arXiv191108450L} compiled a UFD sample providing chemical analysis and detailed description of three UFDs: Bo\"otes~I, II, and Canes Venatici~I. They consider three observational constraints for their models, the present-day stellar mass, the [$\alpha$/Fe]--[Fe/H] relation and the stellar metallicity distribution function (MDF). 
They follow the chemical evolution code of \citet{2003MNRAS.345...71L} with few modifications and run two main branches of models assuming different gwIMFs. One branch assumes the \citet{1955ApJ...121..161S} IMF as the gwIMF and the other the \citet[hereafter IGIMF-R14]{2014MNRAS.437..994R} gwIMF formulation, which is an implementation of the IGIMF theory of \citet{2003ApJ...598.1076K} based on empirical constraints on the IMF variation from a dynamical study of resolved star clusters and dwarf galaxies \citep{2012MNRAS.422.2246M}.
In order to explore input parameter space, \cite{2019arXiv191108450L} assume three different star-formation-rate--instantaneous-gas-mass ratios: $\nu=SFR/M_{\rm g}=$ 0.005, 0.01 and 0.1 [$\mathrm{Gyr}^{-1}$] 
\footnote{From hereon and in compliance with 
\cite{2009ApJ...706..516P}, we refer to the gas-depletion timescale, $\tau_{\rm g} = 1/\nu$, and not the star formation efficiency $\nu$.}
and two total infall gas mass values: $M_{infall}=$ $10^7 M_\odot$ and $2.5 \cdot 10^7 M_\odot$ and fix the dark matter halo mass and effective radius values as constrained by observations. Based on this study the authors conclude that the data are well reproduced assuming the Salpeter gwIMF, and that within the explored parameters, the IGIMF-R14 formulation cannot reproduce the main chemical properties. 

The premise of being able to place constraints on the small-scale empirical prescriptions entering the IGIMF formalism, deserves further research.   
For this purpose we use the publicly available chemical evolution code GalIMF \citep{2019A&A...629A..93Y}, the modular structure of which allows to readily test different empirical prescriptions entering the IGIMF theory therewith allowing to study the chemical properties of UFDs in more detail.

The paper is organized as follows. Section~\ref{sec: Data} introduces the observational constraints. In Section~\ref{sec: Model} we describe the IGIMF theory and galaxy chemical evolution model. The mechanism of how different input parameters affect the model result is explained in Section~\ref{sec: Input parameters}, where the applied input parameters of the best-fit model are also listed. The calculation results are shown in Section~\ref{sec: Results} and compared with the result given by \citet{2019arXiv191108450L}. Then we discuss the robustness of our study and give conclusions in Section~\ref{sec: Discussion} and \ref{sec: Conclusion}, respectively.

\section{Data}\label{sec: Data}

In order to compare with the previous work, we consider the same data set for galaxy mass, mean metallicity, and the metal abundance of single stars compiled in \citet{2019arXiv191108450L}.

However, we consider only Bo\"otes~I instead of all the other galaxies studied by \citet{2019arXiv191108450L} for the following reasons. (i) Unlike Bo\"otes~I (see Section~\ref{sec: Star formation history} below), most galaxies have a complex gas flow and SFH. For example, \citet{2014ApJ...789..147W}, utilizing the colour-magnitude diagram (CMD), suggest that the UFD galaxy Canes~Venatici~I has two separated starbursts.
However, both this study and \citet{2019arXiv191108450L} assume a single and short gas-infall period followed by a single starburst which cannot properly describe galaxies with a complex SFH.
(ii) the galaxies other than Bo\"otes~I have only a few data points available for the [$\alpha$/Fe]--[Fe/H] relation such that the chemical evolution modelling cannot draw a conclusive result. In \citet{2019arXiv191108450L}, there are in total 27 data points for the [Si/Fe]--[Fe/H], [Mg/Fe]--[Fe/H], and [Ca/Fe]--[Fe/H] relations for Bo\"otes~I, while there are only 8 for Bo\"otes~II, 4 for Canes Venatici~I and less than 10 data points for each of the other UFDs.
As a result, the data points of the UFDs other than Bo\"otes~I are not distributed well in the [$\alpha$/Fe]--[Fe/H] plane to make a good comparison with the predicted chemical evolution track. A follow-up study looking into these individual galaxies with a more detailed gas-flow and SFH modelling and better data will definitely be valuable, which we aim to come back to in the future.

In addition to the observational constraints applied in \citet{2019arXiv191108450L}, we compare the gas-depletion timescale, the star formation timescale (SFT), and the resulting SNIa rate of the best-fit model with the independent observational estimation from \citet{2009ApJ...706..516P}, \citet{2014ApJ...796...91B}, and \citet{2012PASA...29..447M}, respectively.

\section{Model}\label{sec: Model}

In order to reproduce the observed properties of the UFD Bo\"otes~I, we use the GalIMF code \citep{2019A&A...629A..93Y} with modifications to compute the galaxy mass, energy, and chemical evolution. The code is publicly available at https://github.com/Azeret/galIMF.

The free input parameters of the GalIMF code (that we adjust to find the best fit with the observations) are the two parameters related to the star formation history (SFH) of the galaxy -- its initial gas mass, $M_{\rm ini}$, and its gas-depletion timescale, $\tau_{\rm g}$. While there are three observations to be fitted -- the [$\alpha$/Fe]--[Fe/H] relation, the present-day galaxy stellar mass, and the present-day stellar mean metallicity. Different gwIMF models can be tested according to whether a good fit is possible.

The gwIMF is computed based on two physical parameters, the star formation rate (SFR) and metallicity, via the IGIMF theory (also with the GalIMF code).

Different versions of the IGIMF theory are tested and they are defined by the different empirical constraints of the underlying small-scale star-cluster formation law since the empirical constraints have certain uncertainty especially for the low-mass stars.

In the following sections, we introduce the IGIMF theory and describe our galaxy chemical evolution code.

\subsection{The IGIMF theory}\label{sec: The IGIMF theory}

The stellar populations form in individual star-forming events in dense regions of molecular clouds (i.e. embedded clusters) with the initial masses of the stars being distributed according to the IMF.

The IMF of stars, i.e., with mass between $0.08 M_\odot$ and $150 M_\odot$, is expected to be different in star formation environment with different gas temperature, density, and metallicity \citep{1996ApJ...464..256A,1996ApJ...468..586A,1998MNRAS.301..569L,2003MNRAS.338..817E,2004MNRAS.354..375S,2007MNRAS.381L..40D}. 
However, such a systematic variation is not strong enough to be determined using the observations of the local Universe where we can resolve and count directly the number of stars, especially because the nature of the IMF estimation in star clusters is complicated (due to stellar evolution and dynamical evolution).
Lacking a concrete evidence for the systematic variation of the IMF, it has been assumed, for simplicity, to be universal and invariant \citep{2002Sci...295...82K,2003PASP..115..763C,2010ARA&A..48..339B} with the canonical two-part power-law form defined in \citet{2001MNRAS.322..231K}, being indistinguishable from the \citet{2003PASP..115..763C} IMF (e.g. \citealt[their fig.~8]{2008MNRAS.386..864D}). 
With better constraints or under more extreme physical environment, such as in starburst and high-redshift galaxies, the IMF variation becomes evident nowadays.

For example, in recent years, there is increasing evidence for a dependency of the IMF on the environment. For example, the over-abundance of massive stars in the Galactic centre \citep{2010ApJ...708..834B}, in ultra-compact dwarf galaxies \citep{1999A&AS..134...75H,2009MNRAS.394.1529D,2010MNRAS.403.1054D,2012ApJ...747...72D,2017A&A...608A..53J}, in Galactic globular star clusters \citep{2012MNRAS.422.2246M,2007ApJ...656L..65D}, in GCs in Andromeda \citep{2016ApJ...826...89Z,2017ApJ...839...60H}, in R136 in the Large Magellanic Cloud \citep{2012A&A...547A..23B,2018Sci...359...69S}, and in the massive metal-poor star cluster NGC 796 \citep{2018ApJ...857..132K}. For a more detailed discussion see \citet{2019arXiv191006971K}.

In addition and as stressed by \citet{2013pss5.book..115K} and \citet{2018PASA...35...39H}, the gwIMF of the entire galaxy appears to be different from the IMF in individual star-forming events (embedded clusters, i.e., a gravitationally driven collective process of transformation of the interstellar gaseous matter into stars in molecular-cloud overdensities on a spatial scale of about one pc and within about one Myr, see \citealt{2017A&A...607A.126Y}). This difference between the IMF and the gwIMF is evident in all types of observations, such as dwarf galaxies \citep{2009ApJ...695..765M,2009ApJ...706..599L,Watts2018}, SDSS star-forming galaxies \citep{2008ApJ...675..163H,2011MNRAS.415.1647G}, starburst galaxies \citep{2017MNRAS.470..401R,2018Natur.558..260Z}, and massive elliptical galaxies \citep{1994A&A...288...57M,2010Natur.468..940V,2015ApJ...806L..31M,2018MNRAS.477.3954P}. 

The empirical systematic variation of the gwIMF, summarized in \citet[their fig. 6]{2017A&A...607A.126Y}, can be explained by the IGIMF theory, where the fundamental idea is to calculate the gwIMF by summing all the IMFs of all embedded clusters forming in the galaxy \citep{2003ApJ...598.1076K}. For example, low-SFR galaxies form mostly low-mass star clusters, the sum of which would result in a top-light gwIMF. On the other hand, a higher galactic SFR generally leads to a more top-heavy gwIMF.

The default formulation we apply for this study to calculate the gwIMF as a function of SFR and gas-phase metallicity is from \citet[here referred to as IGIMF-R14]{2014MNRAS.437..994R} to be comparable with \citet{2019arXiv191108450L}.
At low SFR and high metallicity, the gwIMF becomes top-light\footnote{A top-light IMF contains fewer massive stars than the canonical IMF. The definition of the terms top/bottom-light/heavy is clarified in, e.g., \citet{2018A&A...620A..39J}.} and the bottom part of the IMF (i.e. the IMF for low-mass stars, hereafter low-mass IMF) remains unchanged according to the IGIMF-R14 formulation (see e.g. fig. 1 and 2 of \citealt{2019arXiv191108450L}). 

The detailed IGIMF-R14 formulation has been stated in \citet{2019arXiv191108450L} which we repeat here. 

The embedded cluster mass function (ECMF, $\xi_{\mathrm{ecl}}$) is adopted as a single power-law function with a fixed power-index of 2 (\citealt{2003ARA&A..41...57L,2015A&A...582A..93S} and references therein).
\begin{equation}\label{eq:xi_ecl}
\begin{split}
\xi_{\mathrm{ecl}}(M_{\rm ecl},SFR)=\mathrm{d} N_{\rm ecl}/\mathrm{d} M_{\rm ecl}=
~~~~~~~~~~~~~~~~~~~~~~~~~~~~~~~~~ &\\
\begin{cases}
0, & M_{\rm ecl}<M_{\mathrm{ecl,min}},\\
k_{\mathrm{ecl}} M_{\rm ecl}^{-2}, & M_{\mathrm{ecl,min}} \leqslant M_{\rm ecl}<M_{\mathrm{ecl,max}}(SFR),\\
0, & M_{\mathrm{ecl,max}}(SFR) \leqslant M_{\rm ecl},
\end{cases}
\end{split}
\end{equation}
where $M_{\rm ecl}$ is the stellar mass of an embedded cluster and $k_{\mathrm{ecl}}$ is the normalization factor accounting for the total mass of the embedded clusters formed in the galaxy in a single star formation epoch of about 10 Myr. The lower mass limit of the ECMF, $M_{\rm ecl,min}= 5 M_\odot$, is about the mass of the smallest stellar groups observed \citep{2003MNRAS.346..343K,2012ApJ...745..131K}. The upper mass limit of the ECMF is calculated as a function of the instantaneous SFR, $SFR$:
\begin{equation}\label{eq: Meclmax}
    \mathrm{log}_{10}M_{\rm ecl,max}=A+B\cdot \mathrm{log}_{10}SFR(t),
\end{equation}
where $A = 4.83$ and $B = 0.75$. This formulation has been determined by \citet{2004MNRAS.350.1503W} to be consistent with the observed extragalactic $M_{\rm ecl,max}-SFR$ relation (see also \citealt{2013ApJ...775L..38R}) where the parameter values are provided with uncertainties. We note in addition that the newest IGIMF formulation developed in \citet{2015A&A...582A..93S} and \citet[their eq. 11, 12, and fig. 2]{2017A&A...607A.126Y} has moved to a more generalized formulation without the need to introduce the parameter $A$ and $B$. However, since Eq.~\ref{eq: Meclmax} it is used by \citet{2019arXiv191108450L} we implement it as well for the purpose of an unbiased comparison of our results.

We note that $SFR \lessapprox 3.2 \cdot 10^{-6} M_\odot$/yr leads to a numerical inconsistency because $M_{\rm ecl,max}$ would be smaller than $M_{\rm ecl,min}$ according to Eq.~\ref{eq: Meclmax}. Thus, we cannot reproduce the SFHs of \citet{2019arXiv191108450L} where smaller values of the SFR are reached (the same is for \citealt{2015MNRAS.449.1327V}). \citet{2019arXiv191108450L} solve this inconsistency by fixing $M_{\rm ecl,min}$ to be $5 M_\odot$ at low SFRs (Matteucci, private communication).

Here we decide, however, to simply set the SFR to zero if the calculated $SFR < 3.5 \cdot 10^{-6} M_\odot$/yr.
Because the number of stars and star clusters formed in less than\footnote{The typical timescale for a star cluster to form is $\approx 1$ Myr.} 10 Myr is so small that the IMF can no longer be treated as a smooth and continuous function.
We point the reader to the discussion of extremely low SFRs with $SFR \lessapprox 10^{-6} M_\odot$/yr, the associated phenomenon of H$\alpha$--dark-star-formation, and the associated maximum stellar masses that can form as provided in sec.4 of \citet{2007ApJ...671.1550P}. At extremely low SFRs, the formation of individual stars will be affecting the observational tracers. The IGIMF theory can account for this, as instead of using the smooth gwIMF in integrated form, the formation of individual stars can also be traced \citep[OSGIMF module of the GalIMF code]{2017A&A...607A.126Y}.

The different treatments between our calculation and \citet{2019arXiv191108450L} should not have a significant effect on the result as it only relates to a small fraction of stars. To be specific, the suggested SFT of Bo\"otes~I is less than 1 Gyr. Thus, the largest error on the modeled mass made by neglecting low-SFR activities is $\Delta M = SFR \cdot SFT < 3.5 \cdot 10^{-6} M_\odot/\mathrm{yr} \cdot 10^9 \mathrm{yr} < 3.5 \cdot 10^{3} M_\odot$, which is similar to the observational uncertainty of the mass of Bo\"otes~I (see Table~\ref{tab:parameter}). The metal yield of these low-SFR populations is more negligible as they are composed exclusively by stars with a lower mass ($\lessapprox 3 M_\odot$ according to the OSGIMF module of the GalIMF code) and having a longer lifetime ($\gtrapprox 400$ Myr).


The IMF of stars in an embedded cluster, $\xi_{\star}$, is assumed to be a two-part power-law function with its slope a function of metallicity and its upper mass limit a function of $M_{\rm ecl}$:
\begin{equation}\label{eq:xi_star}
\begin{split}
\xi_{\star}(m,M_{\rm ecl},\mathrm{[Fe/H]}) = \mathrm{d} N_{\star}/\mathrm{d} m= ~~~~~~~~~~~~~~~~~~~~~~~~~~~~~~~~~~~~~~~ &\\
\begin{cases} 
0, & m<0.08~M_{\odot}, \\
k_{\mathrm{\star}} (m/0.5~M_{\odot})^{-1.3}, & 0.08~M_{\odot} \leqslant m<0.5~M_{\odot}, \\ 
k_{\mathrm{\star}} (m/0.5~M_{\odot})^{-\alpha(\mathrm{[Fe/H]})}, & 0.5~M_{\odot} \leqslant m<m_{\mathrm{max}}(M_{\rm ecl}), \\
0, & m_{\mathrm{max}}(M_{\rm ecl}) \leqslant m,
\end{cases}
\end{split}
\end{equation}
where $m$ is the stellar mass and $k_{\mathrm{\star}}$ the normalization factor. The power-law index $\alpha$ is calculated as:
\begin{equation}\label{eq: alpha2}
    \alpha = 2.3 + 0.0572 \cdot \mathrm{[Fe/H]}.
\end{equation}

We note that Eq.~\ref{eq: alpha2} (used by \citealt{2014MNRAS.437..994R} and \citealt{2019arXiv191108450L}) differs from the original \citet{2012MNRAS.422.2246M} formulation which depends additionally on density and only for stars more massive than $1 M_\odot$. 

The calculation of the upper stellar mass limit, $m_{\mathrm{max}}$, in Eq.~\ref{eq:xi_star} is not mentioned clearly in \citet{2019arXiv191108450L} but we assume they follow eq. 3 and 4 of \citet{2009A&A...499..711R} and so do we.

Finally, the gwIMF, $\xi_{\mathrm{IGIMF}}$, is the integration of the IMF of all the embedded clusters that form in the time interval $\delta t=10$ Myr:
\begin{equation}\label{eq:xi_IGIMF}
\xi_{\mathrm{IGIMF}}(m,SFR)=\int_{0}^{+\infty} \xi_{\star}(m,M_{\rm ecl})~\xi_{\mathrm{ecl}}(M_{\rm ecl},SFR)\,\mathrm{d}M_{\rm ecl}.
\end{equation}

\subsection{The IGIMF theory as a framework}\label{sec: The IGIMF theory as a framework}

The IGIMF theory is not a specific IMF formulation but a framework (the mathematical procedure described by Eq.~\ref{eq:xi_IGIMF}) that allows the computation of the gwIMF based on the empirically constrained $\xi_{\star}$ (the constraints are given in e.g. \citealt{2005ESASP.576..629K,2012MNRAS.422.2246M,2019MNRAS.tmp.2759W}) and $\xi_{\mathrm{ecl}}$ \citep{2003ARA&A..41...57L}. Since there is still an uncertainty of how the IMF in individual embedded clusters varies and also of the mass distribution function of the embedded star clusters, the IGIMF theory (Eq.~\ref{eq:xi_IGIMF}) can lead to different gwIMFs depending on the assumptions made for $\xi_{\star}$ and $\xi_{\mathrm{ecl}}$.

This does not mean that one can adjust the IMF (ECMF) formulation to fit with any observation. On the contrary, the independent IMF constraints for low- and high-mass stars (e.g. the observational studies of the IMF in the Milky Way and the dwarf-to-giant ratio sensitive spectral features and UV/H$\alpha$ luminosity ratio of other galaxies) need to be fulfilled and in addition, the resulting gwIMF calculated by the IGIMF theory has to be consistent with the observed galactic mean metallicity in the chemical evolution model where the SFH is determined by other observational constraints. For example, if the gwIMF assumption is more top-light, it has to be also more bottom-light in order to maintain the same resulting mean stellar metallicity, as is discussed in Section~\ref{sec: Mean stellar [Fe/H]} and \ref{sec: Other IGIMF formulations} below.

Three different IGIMF formulations (different assumptions on $\xi_{\star}$ and $\xi_{\mathrm{ecl}}$) are summarized in \citet{2018A&A...620A..39J}: their IGIMF1 formulation does not assume any IMF variation such that the gwIMF variation is purely due to the IGIMF theory; their IGIMF2 formulation considers that the IMF of massive stars depends on the cloud core density and metallicity; and their IGIMF3 model considers in addition that the IMF of low-mass stars depends on the gas metallicity (as supported by empirical evidence such as the one provided by \citealt{2015ApJ...806L..31M}). Different from Eq.~\ref{eq:xi_star}, the assumed power-law $\xi_{\star}$ in \citet{2018A&A...620A..39J} changes at not only $0.5 M_\odot$ but also at $1 M_\odot$, thus having the form
\begin{equation}
\xi_{\star} (m) =    \left\{ \begin{array}{ll}
k_1 m^{-\alpha_1}, \hspace{1.65cm} 0.08\leq m/M_{\odot}<0.50 \,, \\
k_2 m^{-\alpha_2}, \hspace{1.65cm} 0.50\leq m/M_{\odot}<1.00 \,, \\
k_2 m^{-\alpha_3}, \hspace{1.65cm} 1.00\leq m/M_{\odot}< m_{\mathrm{max}} \,, \\
\end{array} \right.
\label{eq:IMF18}
\end{equation}
with
\begin{equation}\label{eq: alpha18}
\begin{split}
    &\alpha_1=1.3+0.5\cdot [Z],\\
    &\alpha_2=2.3+0.5\cdot [Z],\\
    &\alpha_3=\alpha_3(\mathrm{[Z]}, M_{\rm ecl}),
\end{split}
\end{equation}
where $[Z]=\mathrm{log}_{10}$(Z/Z$_\odot$) and $Z$ is the metal mass fraction of the star-forming molecular cloud. That is, $\alpha_1$ and $\alpha_2$ are functions of the metallicity while $\alpha_3$ is a function of both metallicity and the initial stellar mass of a star cluster (see \citealt{2018A&A...620A..39J} their eq. 6 and 9 for the exact formulation of $\alpha_3$). In addition, \citet{2018A&A...620A..39J} assumes for all their model a $\xi_{\mathrm{ecl}}$ with a variable slope depending on the galaxy-wide SFR (see \citealt{2013MNRAS.436.3309W}, \citealt{2017A&A...607A.126Y}, and \citealt[their eq. 2]{2018A&A...620A..39J}).

The IGIMF-R14 formulation leads to a gwIMF that is in between the IGIMF1 and IGIMF2 formulations since the IMF and ECMF variation exists in IGIMF-R14 but is much weaker than in the IGIMF2 formulation. This work mainly demonstrates the results of the default IGIMF-R14 model but the more recent IGIMF formulations published in \citet{2017A&A...607A.126Y} and \citet{2018A&A...620A..39J} are also tested in Section~\ref{sec: Other IGIMF formulations}.

Note that here we are using the notation IGIMF(Ai, i=1,2,...) where IGIMF(A1)=IGIMF1, IGIMF(A2)=IGIMF2, IGIMF(A3)=IGIMF3 in the previous notation introduced in \citet{2018A&A...620A..39J}. This new notation represents the fact that the IGIMF theory, i.e., the gwIMF, is an integration of all the embedded-star-cluster IMFs \citep{2003ApJ...598.1076K}, remaining unchanged but acting on different assumptions, Ai (i=1, 2, 3), on how the embedded-cluster IMF varies, that is, on the mathematical definition of how the IMF in embedded star clusters depends on the metallicity and density of the star-forming molecular cloud core. 

Every one of these Ai formulations is consistent with the star-formation observed in the Milky Way. They only differ in the mathematical description under extreme star-formation environments which are not well constrained to date. Thus, for example, IGIMF(A3) allows for the IMF to vary with metallicity for massive stars as well as for low-mass stars. 

Furthermore, in Section~\ref{sec: Other IGIMF formulations} below, we use the results of this new study on UFD Bo\"otes~I to provide new constraints on how the IMF for low-mass stars varies at low-metallicity. These new constraints are presented as assumption A4 (Eq.~\ref{eq: alpha}).
We emphasise that we are not adjusting the general IGIMF theory to force improve agreement with the data, but that we use the galaxy-wide data of the UFD Bo\"otes~I, which represents an extreme star-formation environment, to improve our knowledge on pc-scale star formation in this environment. Whether this A4 formulation is correct will be testable using other UFDs subject to the constraint that the formation history of these may differ though.

\subsection{Galaxy chemical evolution model}\label{sec: Galaxy chemical evolution model}

The chemical evolution of the dwarf galaxy is calculated using the galaxy chemical evolution model described in \citet{2019A&A...629A..93Y} applying most of the same assumptions as \citet{2019arXiv191108450L}. Here we mention the model modifications and initial settings for this particular study and refer the reader to \citet{2019A&A...629A..93Y} for a detailed and comprehensive description of the galaxy chemical evolution model.

The SFR at a given timestep is determined by the instantaneous gas mass of the galaxy following a linear relation \citep[their eq. 10]{2019arXiv191108450L}. This leads to basically a constant SFR over time before the onset of the galactic wind or any other mechanism that removes the galactic gas efficiently since the change in stellar mass is much smaller than the gas mass.

A simplification of our model is that all gas exists in the galaxy at the initial time while \citet{2019arXiv191108450L} applied a short gas infall timescale. These treatments are equivalent to each other in this particular case since the suggested gas-infall timescale is only 5 Myr, smaller than the smallest time step of our model, $\delta t=10$ Myr.

We adopt the stellar yield of massive stars from \citet{2006ApJ...653.1145K}. However, we only adopt the yield table for metallicity $Z$=0.02, 0.004, and 0. For the stars with an initial metallicity above 0.02, we apply the $Z$=0.02 yield table, while for stars with an initial metallicity below 0.02 we use an interpolated value\footnote{This is different from \citet{2019A&A...629A..93Y} where the yields are only interpolated as a function of stellar mass instead of being interpolated both as a function of stellar mass and initial metallicity.}. For the low- and intermediate-mass stars, our applied yield table from \citet{2001A&A...370..194M} is different from the one adopted by \citet{2019arXiv191108450L}. But since the SFT of our best-fit model is short, the low-mass stars do not have a chance to participate in the chemical evolution of the galaxy.

The contribution to metal enrichment by SNIa was included in galactic chemical evolution models for the first time by \citet{1986A&A...154..279M} with a numerical formulation. We adopt the SNIa yield table from \citet[their W70 model]{1999ApJS..125..439I}. 

The explosion of SNIa follows certain delay time distribution functions (DTD). An analytical DTD was first introduced in \citet{2005A&A...441.1055G}. Here we test two different DTDs, one being an empirical function and another being a physically motivated function.

The empirical power-law DTD suggested by \citet[see their eq. 13 and fig. 7]{2012PASA...29..447M} is applied as our default model. \citet[their equation 3]{2019A&A...629A..93Y} follows basically the same formulation and normalization parameter, with which the SNIa rate peaks at 40 Myr after a single stellar population forms such that the DTD for all stellar population peaks at about SFT plus 40 Myr. The resulting DTD peak in our model is earlier than that of \citet{2019arXiv191108450L} as is shown in Fig.~\ref{fig:SNIa_3}. This difference does not affect our conclusions (see below).

In addition, the solution with the physically motivated DTD applied by \citet{2019arXiv191108450L} is discussed in Section~\ref{sec: The DTD of SNIa events}. Considering that the shape of the real DTD is likely to have a little plateau at early times according to the single-degenerate or double-degenerate SNIa model (see e.g. \citealt{2009A&A...501..531M}), it is necessary to test the single-degenerate DTD formulation given by \citet[their eq. 2]{2001ApJ...558..351M} following \citet{2019arXiv191108450L}.

In both of these DTD models, the total number of SNIa events from a given stellar population not only depends on the mass of the stellar population but also on its IMF. The gwIMF is no longer a power-law when the IGIMF theory is applied and the number of potential SNIa precursors is different to that in a canonical IMF. The calculation of the SNIa number is important and needs to be documented clearly in the models applying a non-canonical IMF. 

For the power-law DTD, a correction to the number of SNIa events is made according to the number of stars with a mass between 1.5 and 8 $M_\odot$ as explained in \citet[their eq. 4]{2019A&A...629A..93Y}.

For the single-degenerate DTD, we apply that of \citet[their eq. 2]{2001ApJ...558..351M} where the mass function therein, $\phi$, is the gwIMF calculated by the IGIMF theory.

A (phantom\footnote{Since the existence of dark matter particles remains speculative \citep{2012PASA...29..395K,2015CaJPh..93..169K}, we refer to the phantom dark matter which is the Milgromian gravitational force \citep{2012LRR....15...10F,2013MNRAS.432.2846L,2015CaJPh..93..232L} but parameterised here, for the sake of comparison with \citet{2019arXiv191108450L}, the same way as dark matter.}) dark matter halo mass of $3 \cdot 10^6 M_\odot$ \citep{2014ApJ...783....7C}, an effective radius of the luminous (baryonic) component of 242 pc \citep{2008ApJ...684.1075M}, and a ratio between the half-light radius and the (phantom) dark matter halo effective radius of 0.3 \citep{2019arXiv191108450L} is assumed. 

With a given initial baryonic mass, a nominal gas binding energy can be estimated using its current-day effective radius, current-day (phantom) dark matter halo, and applying the formulations given in \citet{1998A&A...337..338B}. 

We note, however, that by assuming the estimated (phantom) dark matter mass as an invariant, the gas dominates the galaxy mass initially since the assumed initial/infall gas mass of our or \citet{2019arXiv191108450L}'s best-fit solution is about $4 \cdot 10^6 M_\odot$ or $>10^7 M_\odot$, respectively. It is possible that the galaxy losses most of its gas and also its dark matter halo due to tidal stripping such that the applied gas binding energy is underestimated. This tidal stripping scenario is no longer consistent with the galactic wind assumption (see paragraphs in below and Section~\ref{sec: Wind efficiency}) and is beyond the scope of a chemical evolution model that gives constraints only to the baryonic matter and SFH.

The nominal binding energy is then compared with the total energy deposited in the gas by the type II supernova (SNII) and SNIa events to determine whether a galactic wind develops. Since the gas binding energy changes insignificantly due to the mass transfer between gas and stars compared to the energy generated by the stars, we consider the nominal binding energy a constant for the current purpose and calculate only the initial gas binding energy.

Every star above 8 $M_\odot$ is assumed to explode as SNII at the end of their life. Each SNII and SNIa is assumed to pass a kinetic energy of $\eta_{\rm SNII}\cdot10^{51}$ erg and $\eta_{\rm SNIa}\cdot10^{51}$ erg to the gas-phase, respectively. Following \citet{1998A&A...337..338B} and \citet{2011A&A...531A.136Y}, the thermalization efficiency of SNII and SNIa are $\eta_{\rm SNII}=0.03$ and $\eta_{\rm SNIa}=0.8$, respectively. The parameter $\eta_{\rm SNII}$ is possibly in between 0.01 and 0.1 \citep{2015MNRAS.446.4220R}, while the parameter $\eta_{\rm SNIa}$ can be higher since they occur at a later time in a hotter and more rarefied medium \citep{2011A&A...531A.136Y}. 


The galactic wind model is implemented according to  \citet[their eq. 12]{2019arXiv191108450L}. In this model the galactic mass loss is given by the SFR multiplied by the wind efficiency factor $\omega$. The galactic wind is launched when the accumulated energy deposited by the supernovae exceeds the nominal galactic binding energy. Once this happens the uniformly mixed gas starts to be removed at a rate of $\omega\cdot SFR$ from the model at each time-step. The energy deposited by stellar winds is about two orders of magnitudes smaller than the energy deposited by the supernova \citep{1998A&A...337..338B, 2015MNRAS.446.4220R} and thus can be safely neglected.

We note that the assumed galactic wind behaviour is a very simplified model. As is discussed in Section~\ref{sec: Wind efficiency} and \ref{sec: Star formation history}, the gas removal from UFDs is probably due to environmental effects in addition to stellar feedback \citep{2015MNRAS.446.4220R,2019A&A...630A.140R}. Thus, $\omega$ is a mathematical simplification used to formulate any possible physical mechanism that leads to the gas removal of UFD Bo\"otes~I and the consequential quenching of the star formation. In addition, $\omega$ is not a free parameter in our model. The only purpose of $\omega$ or the only criteria we apply to determine its value is that our code does reproduce the SFH presented in \citet[their fig. 3]{2019arXiv191108450L} under the same set of assumptions.

\section{Input parameters}\label{sec: Input parameters}

The present models use two free parameters, $\tau_{\rm g}$ and $M_{\rm ini}$, to fit the [$\alpha$/Fe]--[Fe/H] relations and final galaxy mass, respectively, following these steps:

\begin{itemize}
    \item $\tau_{\rm g}$: The gas-depletion timescale determines the SFT, that is, how spread out in time the SFH is for a fixed total stellar mass formed. Given the shape of the top part of the IMF and the SFT, the shape of the [$\alpha$/Fe]--[Fe/H] evolution tracks (i.e. the [Fe/H] value when SNIa start to have a significant influence on [$\alpha$/Fe]) is determined as has been mentioned in \citet{2019arXiv191108450L}. For a given observed [$\alpha$/Fe]--[Fe/H] relation, a more top-light gwIMF requires a shorter $\tau_{\rm g}$ (i.e., larger $\nu=1/\tau_{\rm g}$ as defined in \citealt{2019arXiv191108450L}), to fit the relation.

    A short $\tau_{\rm g}$ is not unprecedented for dwarf galaxy studies (cf. \citealt{2006A&A...453...67L,2015MNRAS.449.1327V}). The conclusion that UFDs have a much longer $\tau_{\rm g}$ than dwarfs stems from the chemical evolution models assuming the Salpeter gwIMF \citep{2014MNRAS.441.2815V,2015MNRAS.446.4220R}, but this conclusion is IMF dependent.

    The parameter $\tau_{\rm g}$ can only fit the average [$\alpha$/Fe]--[Fe/H] relation for the $\alpha$ elements, while the individual [$\alpha$/Fe]--[Fe/H] relations for different $\alpha$ elements depend on the literature stellar yield tables, the SNIa yield, and the DTD of SNIa events.

    \item $M_{\rm ini}$: The galaxy initial mass\footnote{The initial mass is also the total mass since we assume no gas accretion at later times. Thus, $M_{\rm ini}$ is comparable to the parameter $M_{infall}$ of \citet{2019arXiv191108450L}. See Section~\ref{sec: Galaxy chemical evolution model}.} (baryon + phantom dark matter) and radius determines the galactic potential, and thus how many stars are formed until the total energy production from supernovae is equal to the binding energy, which determines the onset time of the galactic wind. With the resulting number of SN for a computed gwIMF, the total stellar mass formed and the final living stellar mass, $M_{\rm *,final}$, of the galaxy is determined. In other words, a higher $M_{\rm ini}$ leads to a deeper galactic potential that allows the formation of more stars before the star formation activity is quenched by the energy feedback.
\end{itemize}

Our best-fit model has a lower $M_{\rm ini}$ and a shorter $\tau_{\rm g}$ compared with the parameters applied in \citet{2019arXiv191108450L} of all their models (their 1BooI to 6BooI). The input parameters are listed in Table~\ref{tab:parameter}.
\begin{table*}
    \caption{Input parameters and computation results of our best-fit chemical evolution models assuming different IGIMF formulations introduced in Section~\ref{sec: The IGIMF theory} (line 3 to 6) compared with the observational values (first line) and \citet{2019arXiv191108450L} (second line). Columns: (1) gwIMF assumption, (2) The assumed DTD model, where PL stands for the power-law DTD given by \citet{2012PASA...29..447M} and SD stands for the single-degenerate DTD given by \citet{2001ApJ...558..351M}, (3) wind efficiency, (4) gas-depletion timescale, and the observational value from an ensemble of dwarf galaxies \citep{2009ApJ...706..516P} in the parentheses, (5) initial gas mass, (6) the starting time of the galactic wind, (7) number of SNIa exploded per 1000 $M_\odot$ of stars formed when the age of the galaxy is 10 Gyr and the observational estimation for a single age stellar population from an ensemble of stellar systems \citep{2012PASA...29..447M} in the parentheses, (8) whether the model-predicted metal enrichment history fits reasonably well with the observed [$\alpha$/Fe]--[Fe/H] relation, (9) total mass of living stars at 14 Gyr and the observational value for Bo\"otes~I \citep{2008ApJ...684.1075M}, (10) stellar-mass-weighted [Fe/H] for the living stars at 14 Gyr and the observational (number averaged) value for Bo\"otes~I \citep{2019arXiv191108450L}. Other parameters are stated in the text. The columns are divided by the vertical lines into five regions, highlighted by the shaded colour. From left to right, the regions indicate: fixed assumptions/parameters of the model, variable parameters to obtain a better fit, results (by-products) of the model that cannot be well-constrained by observation, target parameters that we try to fit, and the additional observable that provides a test to a model.}
    \label{tab:parameter}
    \centering
    \begin{tabular}{bbb|aa|bb|aa|b}
    \hline
    Label & DTD & $\omega$ & $\tau_{\rm g}$ (Gyr) & $M_{\rm ini}$ ($10^{6}$ $M_\odot$) & $t_{\rm wind}$ (Gyr) & $N_{\rm SNIa,gal}$ & [$\alpha$/Fe] & $M_{\rm *,final}$ ($10^{4}$ $M_\odot$) & [Fe/H]$_{\rm mean}$\\ \hline
    Observation\tablefootmark{a} & - & - & $(2.52^{+3.8}_{-2.5})$ & - & - & $(2\pm0.7)$ & - & $3.4\pm0.3$ & $-2.35\pm0.08$\\ 
    Lacchin19\tablefootmark{b} & PL & -\tablefootmark{c} & 10 & 10 & 0.19 & - & N\tablefootmark{d} & 17 & -2.5\\ 
    IGIMF-R14 & PL & 100 & 5.9 & 4.02 & 0.05 & 3.04 & Y\tablefootmark{d} & 3.15 & -2.18\\ 
    IGIMF-R14-SD & SD & 100 & 6.7 & 4.33 & 0.06 & 3.65 & Y\tablefootmark{d} & 3.56 & -2.50\\ 
    IGIMF(A2) & PL & 100 & 2.2 & 2.31 & 0.05 & 1.96 & Y\tablefootmark{d} & 3.66 & -2.70\\ 
    IGIMF(A3) & PL & 100 & 5.0 & 5.63 & 0.06 & 4.87 & Y\tablefootmark{d} & 2.61 & -1.93\\ 
    IGIMF(A4) & PL & 100 & 2.86 & 3.25 & 0.05 & 3.30 & Y\tablefootmark{d} & 3.40 & -2.36\\ 
    \hline
    \end{tabular}
    \tablefoot{
        \tablefoottext{a}{The parentheses indicate that the quantity inside is estimated for an ensemble of stellar systems.}
        \tablefoottext{b}{Here we list only the 3BooI model of \citet{2019arXiv191108450L} that assumes the IGIMF theory. \citet{2019arXiv191108450L} list a grid of model results instead of a best-fit model. Their different models fit different observations well but not all observations simultaneously.}
        \tablefoottext{c}{The $\omega$ stated in \citet{2019arXiv191108450L} should not be applied (see Section~\ref{sec: Wind efficiency} below).}
        \tablefoottext{d}{The "N" stands for "does not fit" while Y stands for "acceptable fit". The resulting [$\alpha$/Fe]--[Fe/H] relation of the 3BooI model assuming IGIMF-R14 in \citet{2019arXiv191108450L} is shown by the dashed line of Fig.~\ref{fig:MgSiCa_3}. The results from our model IGIMF-R14 and IGIMF-R14-SD are shown in Fig.~\ref{fig:MgSiCa_3} and \ref{fig:MgSiCa_5}, respectively. The results from model IGIMF(Ai) are not shown but fit the data as well as model IGIMF-R14.}
    }
\end{table*}

We note that the model suffers from the limited time resolution of the galaxy chemical evolution model when our best-fit SFT is a few 10 Myr while the shortest timestep in our model is $\delta t=10$ Myr. 
That is, all the stars formed within the first 10 Myr timestep have zero metallicity even if some of them would already be enriched in reality.

Since the gwIMF depends on the gas metallicity using the IGIMF-R14 formulation, we must artificially define the logarithm of the metallicity of the first timestep to be $[Z]=-7$ instead of $[Z]=-\infty$. This timestep effectively introduces some uncertainty to our results.

However, we consider the choice of initial metallicity to not be a free parameter of our model because: (1) the galaxy metallicity of the second timestep is about $[Z]=-4$ no matter what initial metallicity is chosen such that the initial metallicity must be lower than $[Z]=-4$; and (2) the IGIMF formulation is supported by empirical evidence for $[Z]\gtrapprox-4$. The resulting gwIMF can no longer be trusted when $[Z]$ is much lower (as discussed in Section~\ref{sec: Other IGIMF formulations} below). Therefore, to mimic zero initial metallicity, we consider $[Z]=-7$ a reasonable choice.

We emphasise the importance of exploring large enough parameter space. While full blind numerical exploration would be computational time-wise very challenging, we present a systematical way of exploring physically plausible values of input parameters. Our recovered input parameters, with which the model fits the observations, are outside of the explored parameter space of \citet{2019arXiv191108450L}.

Due to the above mentioned high computational costs, we do not numerically optimise the model. It turns out that (see the next Section) our model agrees well with the data, thus fulfils the aim of this work -- that is to investigate whether it is possible to reproduce observed properties of Bo\"{o}tes I naturally within the IGIMF framework.

\section{Results}\label{sec: Results}

\begin{figure}
    \centering
    \includegraphics[width=\hsize]{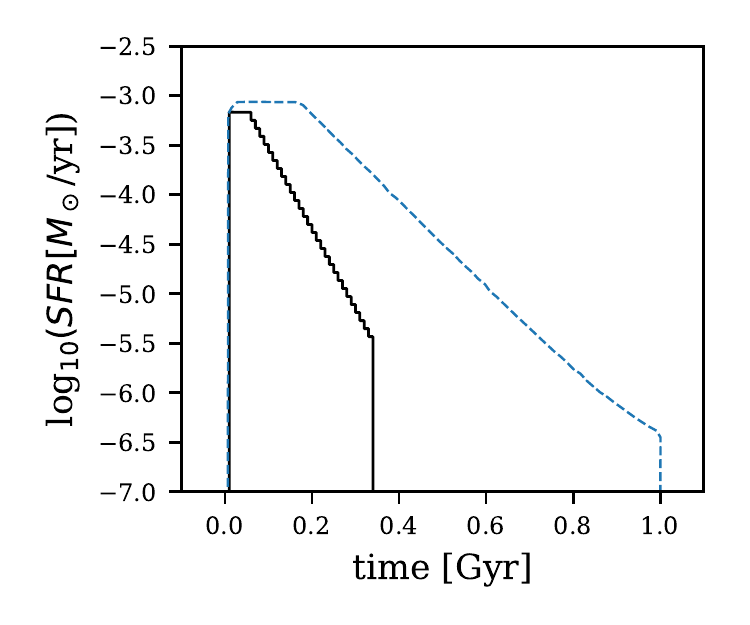}
    \caption{SFHs. The SFH of the best-fit IGIMF-R14 model adopting the parameter set in Table~\ref{tab:parameter} is shown with the black line. The SFH is shown as a histogram because the smallest timestep is 10 Myr. The starting time of the galactic wind is 50 Myr (cf. Fig.~\ref{fig:energy_evolution_3}) while the SFR drops to half of its peak value at about 90 Myr, which defines the SFT. The blue-dashed line is the blue-dashed line in \citet[their fig. 3]{2019arXiv191108450L}, i.e., their 3BooI-IGIMF model.}
    \label{fig:SFH_3}
\end{figure}

\begin{figure}
    \centering
    \includegraphics[width=\hsize]{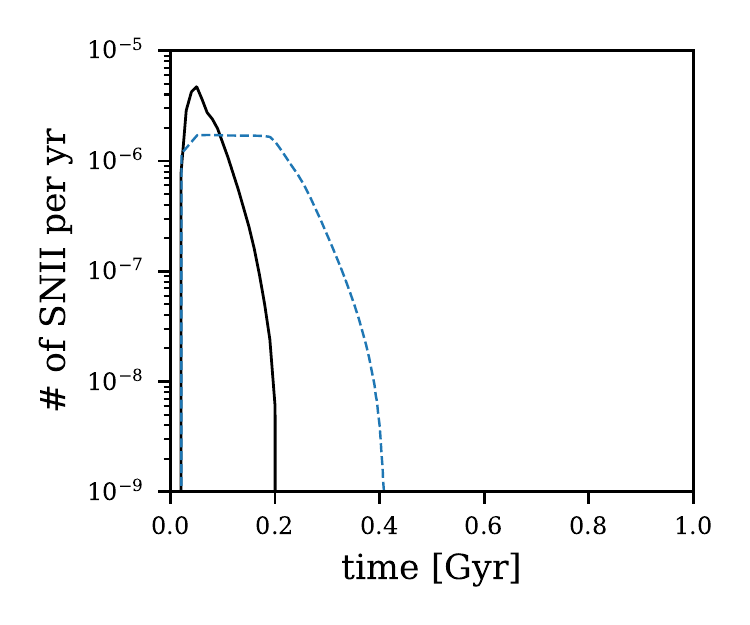}
    \caption{SNII rate evolution history. Similar as Fig.~\ref{fig:SFH_3} but for the SNII rate of the best-fit IGIMF-R14 model. The initial spike SNII rate is due to the discontinuous 10 Myr timestep where the metallicity increases from $[Z]=-7$ to $[Z]=-4$ at the second timestep that leads to a sudden change of the gwIMF according to the IGIMF-R14 formulation (see Fig.~\ref{fig:TIgwIMF_3}). The blue-dashed line is the blue-dashed line in \citet[their fig. 5]{2019arXiv191108450L}, i.e., their 3BooI-IGIMF model.}
    \label{fig:SNII_3}
\end{figure}

\begin{figure}
    \centering
    \includegraphics[width=\hsize]{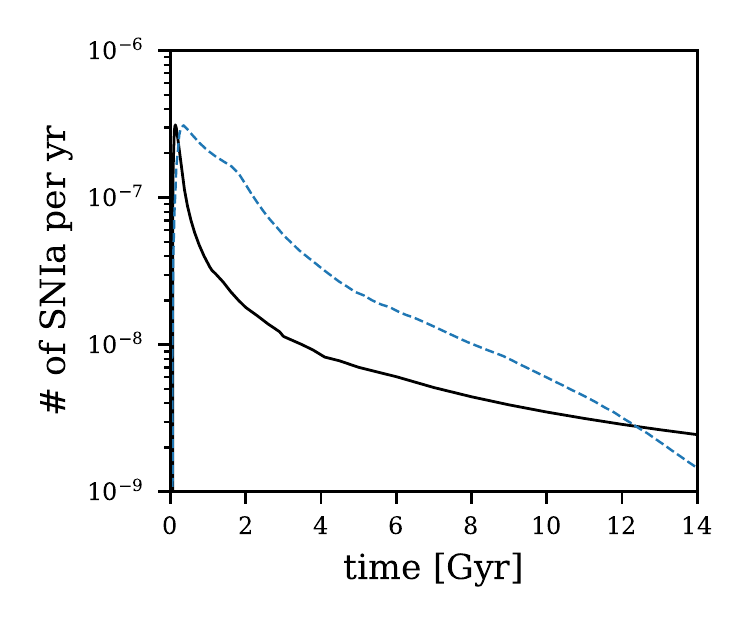}
    \caption{SNIa rate evolution history. Similar to Fig.~\ref{fig:SFH_3} but for the SNIa rate of the best-fit IGIMF-R14 model. The blue-dashed line is the blue-dashed line in \citet[their fig. 4]{2019arXiv191108450L}, i.e., their 3BooI-IGIMF model. We note that the horizontal axis is different from Fig.~\ref{fig:SNII_3}. We test also the DTD assumption of \citet{2019arXiv191108450L} in our IGIMF-R14-SD model, shown in Fig.~\ref{fig:SNIa_5}.}
    \label{fig:SNIa_3}
\end{figure}

\begin{figure}
    \centering
    \includegraphics[width=\hsize]{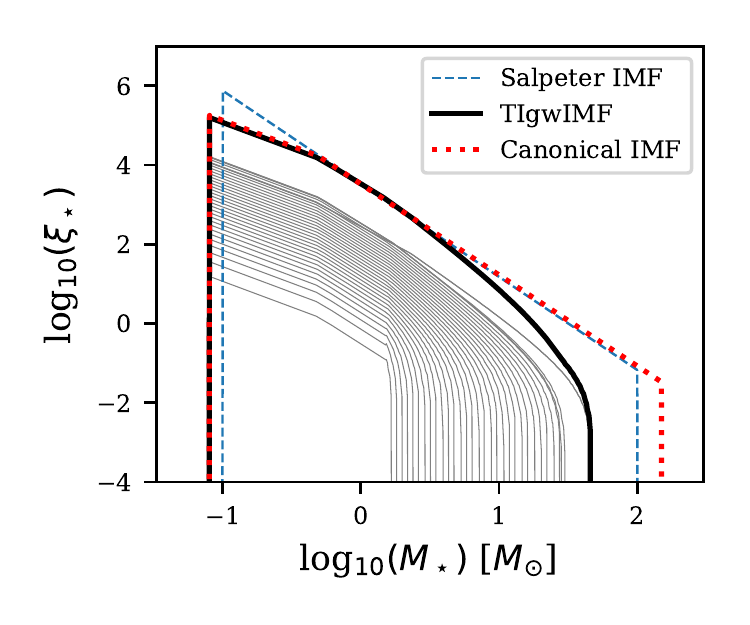}
    \caption{The cumulative (time-integrated) gwIMF (TIgwIMF) for all stars ever formed (the thick solid line) and the gwIMF for each 10 Myr star formation epoch (the thin solid lines, evolving from the top to the bottom as time progresses) of the best-fit IGIMF-R14 model, compared with the canonical IMF and the Salpeter IMF. We note that the gwIMF for the first epoch (the top-most thin gwIMF line) has a significantly higher maximum stellar mass limit than the second epoch. This is only due to the assumption of initial metallicity being Z=$0.02\cdot10^{-7}$ and that the 10 Myr timestep is unresolved.}
    \label{fig:TIgwIMF_3}
\end{figure}

\begin{figure}
    \centering
    \includegraphics[width=\hsize]{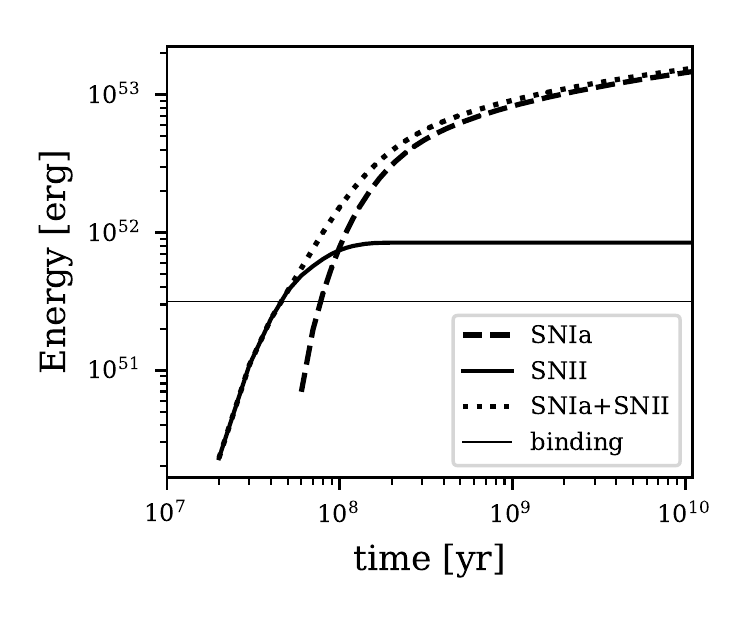}
    \caption{The evolution of energy deposited in the gas by supernovae of the best-fit IGIMF-R14 model, compared to the (initial) nominal gas binding energy assuming that the galactic radius and (phantom) dark matter mass does not evolve significantly (see Section~\ref{sec: Galaxy chemical evolution model}). The galactic wind develops at the timestep after the energy in gas exceeds the nominal binding energy.}
    \label{fig:energy_evolution_3}
\end{figure}

\begin{figure}
    \centering
    \includegraphics[width=\hsize]{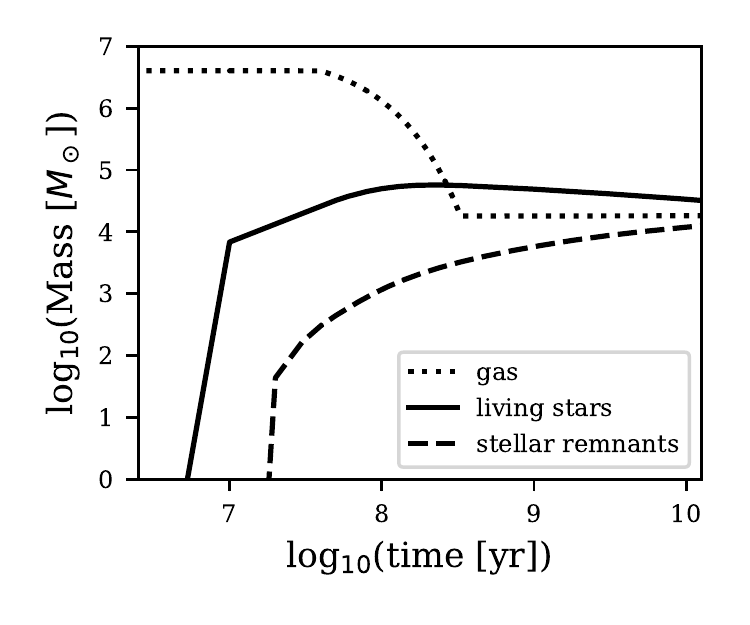}
    \caption{The mass evolution of gas, living stars, and stellar remnants of the best-fit IGIMF-R14 model. The gas mass stops decreasing at about $10^{8.5}$ yr, when the star formation and the corresponding galaxy wind stops, and stays constant thereafter. The amount of remaining gas mass is discussed in Section~\ref{sec: Results}.}
    \label{fig:mass_evolution_3}
\end{figure}

\begin{figure}
    \centering
    \includegraphics[width=\hsize]{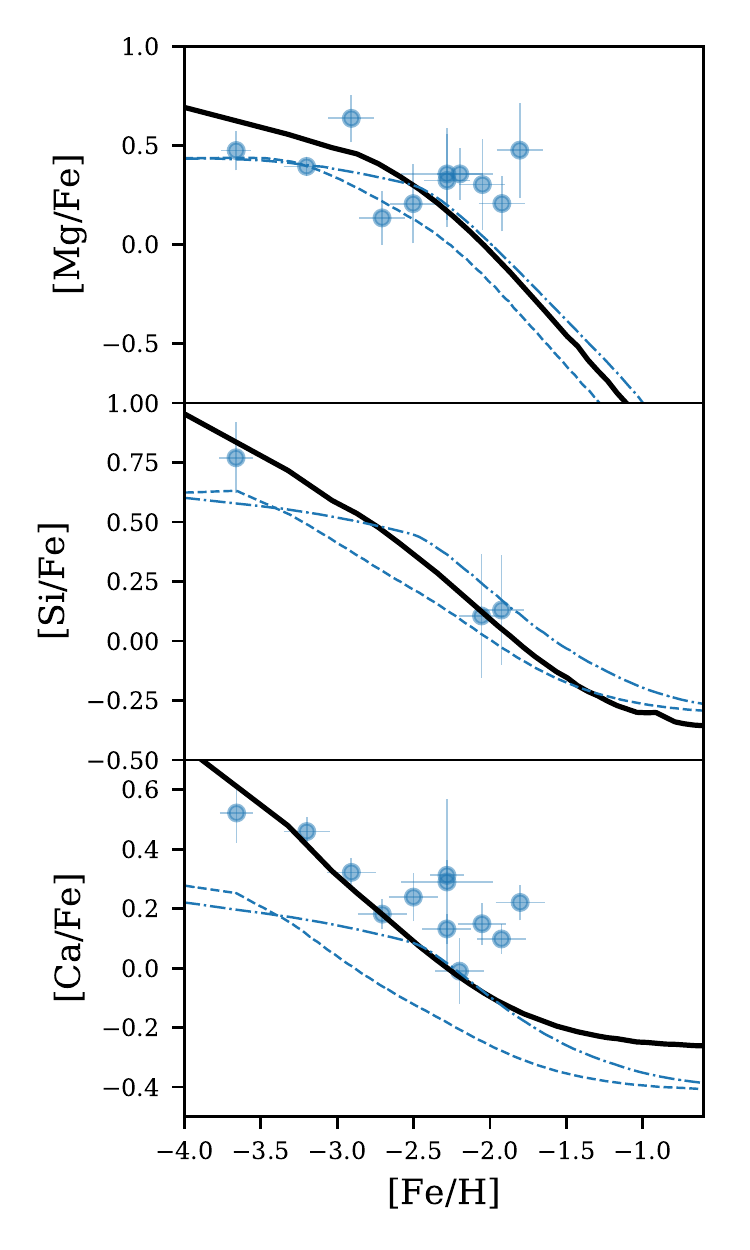}
    \caption{The evolution of [$\alpha$/Fe]--[Fe/H] relations of the best-fit IGIMF-R14 model (black solid line). The circles are the data \textbf{of Bo\"otes~I} collected in \citet[their fig. 6]{2019arXiv191108450L}. The dashed line and the dash-dotted line are the two 3BooI models in \citet[their fig. 6]{2019arXiv191108450L} assuming the gwIMF given by IGIMF-R14 and the Salpeter IMF, respectively.}
    \label{fig:MgSiCa_3}
\end{figure}

A solution of our default model, IGIMF-R14, is calculated following the procedure explained in Section~\ref{sec: Input parameters}. The resulting SFH, SNII and SNIa rates, the accumulated gwIMF of each timestep, and galaxy mass and energy evolution are shown in Fig.~\ref{fig:SFH_3} to \ref{fig:mass_evolution_3}, being compared with the 3BooI-IGIMF model of \citet{2019arXiv191108450L} if a similar plot is provided in their paper.
The resulting [$\alpha$/Fe]--[Fe/H] evolution tracks and final galaxy mass do fit well with the observations as is shown in Fig.~\ref{fig:MgSiCa_3} and Table~\ref{tab:parameter}, respectively. The final galaxy mass lies within a 0.83 $\sigma$ uncertainty range of the observed value. 

The final gas mass shown in Fig.~\ref{fig:mass_evolution_3} represents an upper limit, as the gas mass value is subject to the assumed gas-removal mechanisms that are not taken into account in our computation. In reality, mechanisms such as tidal stripping would additionally remove the gas.

Concerning whether the best-fit $\tau_{\rm g}$ is consistent with the observation of dwarf galaxies, it is important to remember that the observational gas-depletion timescale depends on the gwIMF assumption since both galaxy mass and galaxy SFR estimation depends on the assumed gwIMF. Therefore, we must consider the gas-depletion timescale estimation applying the IGIMF theory. This is given in \citet{2009ApJ...706..516P}. The estimated gas-depletion timescale for star-forming dwarf galaxies becomes shorter when the IGIMF theory is assumed since their gwIMF is expected to be top-light, thus more stars are actually forming per observed H$\alpha$ or UV photon than the estimation assuming the canonical or Salpeter IMF (see Fig.~\ref{fig:TIgwIMF_3} below for a comparison between different IMFs). 

The best-fit $\tau_{\rm g}$ is almost at the centre of the gas-depletion timescale distribution of a set of local star-forming galaxies given in \citet[their fig. 8]{2009ApJ...706..516P} assuming the IGIMF theory, showing that our estimation from chemical abundances is consistent with the estimation from measuring the galactic SFR and gas mass of star-forming galaxies. This self-consistency when applying the IGIMF theory is encouraging. 

The corresponding SFT of the best-fit model also agrees with the morphology of the colour-magnitude diagram (CMD) and the required minimum SFT indicated by the metal distribution of the stars. \citet{2014ApJ...796...91B} show that the two-stellar-component best-fit stellar populations have an age difference of 0.1 Gyr, with one population contributing 97\% of the stellar mass, indicating that Bo\"otes~I is likely to have formed in a single starburst in a short timescale less than 100 Myr, in agreement with \citet{2012ApJ...744...96O} and \citet{2015ApJ...799L..21W}.

Finally, we check if our galactic stellar mean [Fe/H], $\mathrm{[Fe/H]}_{\rm mean}$, agrees with the observation. The result from the best-fit IGIMF-R14 model is listed in Table~\ref{tab:parameter}, which is within a 1.75 $\sigma$ uncertainty range of the observed value. 


Compared to the fitting results assuming the Salpeter gwIMF provided by \citet{2019arXiv191108450L}, our model assuming IGIMF-R14 fits better with the observations. The goodness of the fits for the [$\alpha$/Fe]--[Fe/H] relations is similar between our model and the Salpeter model of \citet{2019arXiv191108450L} as is compared in Fig.~\ref{fig:MgSiCa_3}. However, our $M_{\rm *,final}$ and $\mathrm{[Fe/H]}_{\rm mean}$ fit the observational constraints better than the results given in \citet[their table 3]{2019arXiv191108450L}. 

The IGIMF theory thus leads to a remarkably simple and self-consistent understanding of the Bo\"otes~I UFD galaxy.

\section{Discussion}\label{sec: Discussion}

\subsection{The number of SNIa events}\label{sec: The number of SNIa events}

The total numbers of SNIa events per unit total stellar mass for the modelled Bo\"otes~I UFD galaxy, $N_{\rm SNIa,gal}$, are listed in Table~\ref{tab:parameter}. These values can be compared, although shall not be constrained, by the estimated value from the local universe \citep{2012PASA...29..447M}. $N_{\rm SNIa,gal}$ at a given time is calculated by adding up the SNIa events from each star formation epoch of duration $\delta t = 10$ Myr. The number of SNIa events for each epoch depends on the mass of formed stars and the gwIMF of that epoch.

\citet{2012PASA...29..447M}, assuming an invariant IMF (defined as the "diet-Salpeter IMF"), applied a normalization parameter for the DTD of a single star formation epoch, $N_{\rm SNIa}$, of 2 and 2.2, which is the number of SNIa per 1000 $M_\odot$ of stars formed in 10 Gyr, in their eq. 13 and sec. 4.2, respectively, while \citet{2019A&A...629A..93Y} applied $N_{\rm SNIa}=2.25$.
Here we follow \citet[their sec. 4.2]{2012PASA...29..447M} for the default power-law DTD model, that is $N_{\rm SNIa}=2$.

We note that the observational uncertainty as well as the computational error\footnote{The GalIMF code applies a numerical integration of the IMF to calculate $N_{\rm SNIa}$ which differs from the analytical result with an error of about $\pm3\%$. Numerical integration is necessary when the gwIMF is no longer a power-law function according to the IGIMF theory.} of $N_{\rm SNIa}$ is large (see \citealt[their table 1]{2012PASA...29..447M}) and can lead to a significant difference in the resulting mean stellar [Fe/H]. A one or two sigma difference between the model and observational mean stellar [Fe/H], which is only about 0.1 dex, can be caused by a different normalization parameter of the SNIa as we have tested.

Although the SNIa events affect the MDF differently compared to SNII events such that the [Fe/H] distribution of the observed stars can in principle constrain $N_{\rm SNIa}$, the MDF needs to be subjected to the same observational biases when comparing to the stellar observed one. For example, the galactic abundances are largely based on the measurements of bright giant stars, but these constitute a relatively narrow stellar-age window while the real MDF may have a different mean value, be more spread out, and have a different shape.

Since the observational bias that may affect the shape of the MDF has not been studied in detail, it is not reliable to constrain the $N_{\rm SNIa}$ with the shape of MDF.

\subsection{The DTD of SNIa events}\label{sec: The DTD of SNIa events}

The DTD of SNIa events is uncertain and still under debate. There are two main groups of DTD formulations described by power-law functions and exponential functions, where the power-law DTDs are more peaked at early times (see DTD comparisons e.g. in \citealt{2020arXiv200105967S,2009A&A...501..531M}). Here we test two commonly applied DTD formulations belonging to these two groups, with one of them being the formulation applied by \citet{2019arXiv191108450L}, to explore the potential influence of the DTD assumption on our conclusions.

Our default model, IGIMF-R14, follows the assumptions of \citet{2019arXiv191108450L} closely except that it assumes the power-law DTD described in Section~\ref{sec: Galaxy chemical evolution model} and shown in Fig.~\ref{fig:SNIa_3}. The SNIa rate peaks at about 130 Myr in the IGIMF-R14 model here (which is roughly SFT + 40 Myr) while it peaks at about 350 Myr for the model shown in \citet{{2019arXiv191108450L}}, that is the blue dashed line in Fig.~\ref{fig:SNIa_3}.

The apparent difference in the SNIa event distribution shown in Fig.~\ref{fig:SNIa_3} is not only due to the difference in DTD of the single stellar population but also due to the difference in the SFH (Fig.~\ref{fig:SFH_3}).

The different DTD affects the best-fit SFT since it is the SFT relative to the peak SNIa rate timescale which determines the shape of the [$\alpha$/Fe]--[Fe/H] relations.
Therefore, applying a different DTD would not change the fact that a solution is possible but the corresponding metal-enrichment history would be different.

To ensure a fair comparison and demonstrate the effect of assuming an alternative DTD, we test the DTD applied by \citet{2019arXiv191108450L} in our model IGIMF-R14-SD.

IGIMF-R14-SD applies the single-degenerate DTD formulation of \citet[their eq. 2 to 5]{2001ApJ...558..351M} with the parameters therein $M_{Bm}=3$ and $\gamma=2$. The total number of SNIa is normalized (with the parameter $A$ in \citealt{2001ApJ...558..351M}) such that the GalIMF code reproduces the 3BooI-Salpeter model of \citet{2019arXiv191108450L}, i.e., resulting in the same SFH and SNIa rate evolution history. 

According to the single-degenerate SNIa model assumption \citep{1983A&A...118..217G}, the SNIa rate is determined by the death rate of the companion star with a mass $M_2$ (while the initial mass of the primary star is $M_1$). Thus, to obtain all SNIa events, we need the mass distribution of the companion stars, $\widetilde{\Psi}(M_2)$, and the lifetime function. 
The mass distribution of the companion stars is calculated by integration over the possible total masses of binary systems ($M_{\rm B}=M_1+M_2$) that possess a secondary star with mass $M_2$ modulated by a probability function of having such a binary system given the value $M_2/M_{\rm B}$.

As discussed, for example, in \citet{2018arXiv180610605K}, the IMF of the binary systems (their “the IMF of unresolved binaries”) is similar to the IMF of single stars. Therefore, when calculating the SNIa rate, it is reasonable to use the gwIMF calculated for the single stars at a given time as the IMF of the $M_{\rm B}$.


The calculation results of model IGIMF-R14-SD are shown in Table~\ref{tab:parameter} and Appendix~\ref{sec: other_results}.

The result of model IGIMF-R14 and IGIMF-R14-SD applying two different DTDs both fit the observational values well and within about 2 standard deviations (Fig.~\ref{fig:MgSiCa_3}, \ref{fig:MgSiCa_5}, and Table~\ref{tab:parameter}). 

Notably, the observational value is in between the results of these two models, indicating that the IGIMF-R14 formulation agrees with the data with a reasonable DTD assumption.

\subsection{Galaxy age}\label{sec: Galaxy age}

The galaxy evolution model stops at a certain time, according to the estimated age of Bo\"otes~I, and generates model outputs. The stopping time needs to be specified before simulation begins. Here we set the modelled galaxy to be 14 Gyr old.

Through a CMD fitting analysis, Bo\"otes~I is found to be dominated by an old single stellar population of age $=14\pm2$ Gyr according to \citet{2008AJ....135.1361D}, and of about $13.7$ Gyr according to \citet{2012ApJ...744...96O}, and of $13.3\pm0.3$ Gyr according to \citet{2014ApJ...796...91B}. 

The uncertainty of the age estimation comes from the uncertainty of the isochrone model and the metallicity assumption of the applied isochrone \citep{2020MNRAS.491.2025C}. The metallicity and $\alpha$-enhancement of the real stellar population are not exactly the same as the assumed isochrone.

As the galaxy becomes older, its living stellar mass (Fig.~\ref{fig:mass_evolution_3}) and mean stellar [Fe/H], which we compared with the observed values, decreases slowly. An age difference of about 1 or 2 Gyr does not have a significant effect on the results of the galaxy evolution model.

\subsection{Final mass of living stars}

The mass of living stars at 14 Gyr in our model is compared and fitted with the observational stellar mass of Bo\"otes~I by modifying the input parameter $M_{\rm ini}$ as described in Section~\ref{sec: Input parameters}.

The observational stellar mass of Bo\"otes~I estimated by \citet{2008ApJ...684.1075M} depends on the assumed gwIMF, and in particular on the gwIMF of low-mass stars, since the galaxy in question is about 13.7 Gyr old. 

Here we compare our result with the stellar mass estimation assuming the canonical IMF because the IGIMF-R14 formulation, despite having a systematically changing gwIMF, also assumes a fixed canonical IMF for stars less massive than 1 $M_\odot$. Thus the comparison is consistent.

The low-mass part of the IMF may also change depending on the metallicity of the star-forming region. Such a variation has been suggested by the observation of massive elliptical galaxies (e.g., \citealt{2010Natur.468..940V}, \citealt{2015ApJ...806L..31M}, and \citealt{2018MNRAS.477.3954P}) and already earlier by evidence gleaned from resolved stellar populations \citep{2002Sci...295...82K,2012MNRAS.422.2246M,2018A&A...620A..39J}. The IGIMF theory in its most recent formulation \citep{2018A&A...620A..39J} predicts a bottom-light gwIMF for low-metallicity populations. 

If the real gwIMF is indeed bottom-light, the estimated galaxy mass using the star number counts from \citet{2008ApJ...684.1075M} would be lower, but the galaxy mass that best-fits the [$\alpha$/Fe]--[Fe/H] relation with our model will also be lower, therefore, retaining the consistency between our model and the observational mass.

\subsection{Mean stellar [Fe/H]}\label{sec: Mean stellar [Fe/H]}

In this work, the stellar MDF (and its mean, $\mathrm{[Fe/H]}_{\rm mean}$) is not fitted but predicted by the model. Thus, it can be compared with the observations to test the model. But see also Section~\ref{sec: The number of SNIa events} for a discussion of the bias affecting an observed MDF.

\citet{2019arXiv191108450L} compare their model prediction with the MDF while this work compares the model result only with the $\mathrm{[Fe/H]}_{\rm mean}$ as a representative measure of the full MDF.

Due to the small number of observed stars ($\approx 30$) available to construct the observational MDF, the mean is the most prominent and thus robust feature of the MDF. The second most prominent feature, being the width of the predicted MDF, is not significantly changed relative to the observational width of the MDF under different input parameters (as is shown in \citealt[their fig. 8, 13, and 15]{2019arXiv191108450L}). we argue that no useful information is lost when we compress the MDF into its mean. As a result, a mismatch of the MDF between model prediction and observation should be considered as a single mismatch of the mean stellar [Fe/H] instead of repeated failures for every single star.

We note that such a simplification, although appropriate here, would not be appropriate in a study of galaxies with better constrained observational MDFs. In such a case, the MDF is providing useful information in constraining the galaxy chemical evolution model.

The resulting $\mathrm{[Fe/H]}_{\rm mean}$ of the IGIMF-R14 model agrees well with observation \citep{2019arXiv191108450L} within the 1.75 $\sigma$ error range.

The agreement may be undervalued since the observational estimation of [Fe/H] depends on the assumed [$\alpha$/Fe] (e.g., \citealt{2012ApJ...744...96O}). Different groups give different [Fe/H] estimations, indicating a larger uncertainty of the observation. For example, \citet[their table 5]{2008AJ....135.1361D} estimate that the best-fit [Fe/H] value for the CMD is $-2.2\pm0.2$, which agrees perfectly with our default model IGIMF-R14. On the other hand, \citet{2011ApJ...738...51L} apply low-resolution spectroscopic analysis finding that the mean [Fe/H] value of 25 stars is -2.59 (or -2.64 if assuming the solar metallicity from \citealt{2009ARA&A..47..481A}) which is much lower and may be more challenging to reproduce with our current model assumptions. 

We note that the observational mean [Fe/H] from the above literature is a direct mean of [Fe/H] values inferred for the single stars.
While [Fe/H]$_{\rm mean}$, the stellar iron abundance for the entire galaxy, is the mass-weighted [Fe/H] defined as $\mathrm{[Fe/H]}_{\rm mean}=\mathrm{log}_{10}(M_{\rm Fe}/M_{\rm H})-\mathrm{log}_{10}(M_{\rm Fe,\odot}/M_{\rm H,\odot})$, where $M_{\rm Fe/H}$ is the total mass of iron/hydrogen in all the stars in Bo\"otes~I and $M_{\rm Fe,\odot/H,\odot}$ is the total mass of iron/hydrogen in the Sun. Thus, the observational mean [Fe/H] and [Fe/H]$_{\rm mean}$ are not exactly comparable. However, due to the small number of measured stars, this difference is not usually mentioned (e.g. in \citealt{2019arXiv191108450L}).

It is not trivial that the predicted $\mathrm{[Fe/H]}_{\rm mean}$ of our model agrees with the observation automatically, because the procedure described in Section~\ref{sec: Input parameters} only fits the [$\alpha$/Fe]--[Fe/H] relation and the final living stellar mass. The following factors can all affect the final [Fe/H] significantly:
\begin{itemize}
    \item gwIMF: For the same mass in final living stars, a stellar population with a bottom-light gwIMF produces more metals than the case for the canonical IMF and leads to a metal-rich galaxy.
    \item SFT: A shorter SFT leads to a smaller metal-mass returned to the gas before the end of the star formation era, thus decreasing the stellar metallicity.
    \item Galaxy mass: Given the initial galaxy mass, the gas binding energy determines how many stars can form before the onset of the strong galactic wind, and thus how many metals can be produced by the stars. The mean metallicity of the galaxy is determined and depends on the galaxy mass. Because the binding energy is approximately proportional to the square of the galaxy mass while the energy production from the supernovae is linearly related to the galaxy mass. This leads to a higher metallicity for a more massive galaxy.
\end{itemize}

Although the galaxy mass and SFT have been determined in the fitting procedure described in Section~\ref{sec: Input parameters}, the assumed gwIMF can still affect the final galactic metallicity. Thus the galaxy chemical evolution model can potentially falsify the IMF theory applied. But the IGIMF-R14 formulation turns out to work well.

The fit can be improved further if the gwIMF is slightly more bottom-heavy (than the canonical IMF). This would be the case if the low-mass IMF slope is slightly steeper or if the minimum star cluster mass is slightly lower. 

The IMF constraint for the low-mass stars has a large uncertainty. See, for example, \citet[their eq. 2]{2001MNRAS.322..231K} and Section~\ref{sec: Other IGIMF formulations}. Thus, our result is well consistent with the observation, signalling a fruitful potential of the IGIMF theory.

\subsection{Other IGIMF formulations}\label{sec: Other IGIMF formulations}

Since the assumed IMF function affects the chemical evolution of the UFD Bo\"otes~I, we can use the observed properties of Bo\"otes~I to constrain the IMF formulation, i.e., $\xi_*$ (Eq.~\ref{eq:xi_star} and \ref{eq: alpha2}).
Instead of the default IGIMF-R14 formulation, here we test if the IGIMF(A2) and IGIMF(A3) formulations (introduced in Section~\ref{sec: The IGIMF theory as a framework}, Eq.~\ref{eq:IMF18} and \ref{eq: alpha18}) is consistent with the constraints given by Bo\"otes~I.

The [$\alpha$/Fe]--[Fe/H] relation can be fitted regardless of the applied $\xi_*$ variation assumption.
By adjusting $\tau_{\rm g}$, all the IGIMF formulations can develop a galactic wind at a similar time (Table~\ref{tab:parameter}) thus fit the [$\alpha$/Fe]--[Fe/H] relation. The required values of $\tau_{\rm g}$ are also acceptable and agree well with the observational constraint (Table~\ref{tab:parameter}).

However, not every IGIMF formulation can fit the galaxy mass and metallicity simultaneously, as is explained in Section~\ref{sec: Mean stellar [Fe/H]} above. The best-fit IGIMF formulation appears to be in between the IGIMF(A2) and IGIMF(A3) formulations. Once a modification of the low-mass IMF slope is allowed, that is one more free parameter, all the considered observations can be fitted perfectly. This means that the low-mass IMF formulation is constrained by the high-mass IMF (IMF of the stars with a mass higher than 1 $M_\odot$) and the observation of UFDs. 

The best-fit IGIMF formulation, IGIMF(A4), assumes that the power-law index of the low-mass IMF, $\alpha_1$ and $\alpha_2$ in Eq.~\ref{eq:IMF18}, varies with metallicity according to:
\begin{equation}\label{eq: alpha}
\begin{split}
    \alpha_1=1.3+0.12\cdot [Z],\\
    \alpha_2=2.3+0.12\cdot [Z],
\end{split}
\end{equation}
instead of Eq.~\ref{eq: alpha18}. This variation of $\alpha_1$ and $\alpha_2$ is smaller than the assumption applied in \citet{2012MNRAS.422.2246M}, leading to a mildly bottom-light gwIMF. 

The different multipliers of [$Z$] (0.12 in Eq.~\ref{eq: alpha} while 0.5 in \citealt[their eq. 12]{2012MNRAS.422.2246M}) can agree with each other if the IMF slope depends on the metallicity, $Z$, instead of $[Z]$. The latter is not reasonable when $Z\approx 0$ since the IMF shape varies significantly from $[Z]=-10$ to $[Z]=-100$ even if $Z$ is similar to zero for both cases. 

We propose here a new formulation for the variation of the IMF power-law index for low-mass stars, i.e., $\alpha_1$ for stars with mass smaller than 0.5 $M_\odot$ and $\alpha_2$ for stars with mass between 0.5 and 1 $M_\odot$:
\begin{equation}\label{eq: alpha new}
\begin{split}
    \alpha_1=1.3+\Delta\alpha \cdot (Z-Z_\odot),\\
    \alpha_2=2.3+\Delta\alpha \cdot (Z-Z_\odot).
\end{split}
\end{equation}
where $\Delta\alpha\approx35$ fits the observational $M_{\rm *,final}$ and [Fe/H]$_{\rm mean}$ best. 

With $\Delta\alpha=35$, Eq.~\ref{eq: alpha new} and \ref{eq: alpha} gives the same $\alpha_1$ and $\alpha_2$ values when $[Z]\approx-5.8$ (the case for UFDs) and is consistent with \citet[their eq. 12]{2012MNRAS.422.2246M} when $[Z]\approx-1.3$ (the case for the Galactic GCs, \citealt{2002Sci...295...82K}). That is, the proposed formulation (Eq.~\ref{eq: alpha new}) based on new requirements from the UFDs is similar to the previous formulation (Eq.~\ref{eq: alpha}) for the metal-rich regime such that it naturally fulfils the IMF constraints given by the GCs.

Therefore, with better data at hand, the chemical evolution model of UFDs is capable of providing constraints on IMF variations on the sub-pc (embedded cluster) scale. Our prediction (eq.~\ref{eq: alpha new}) can be compared with the low-mass gwIMF in Local Group dwarf galaxies which will be better constrained with the James Webb Space Telescope \citep{2017MNRAS.468..319E}.

\section{Comparison with Lacchin et al. (2019)}\label{sec: Compare with Lacchin et al. (2019)}

This paper benefits greatly from \citet{2019arXiv191108450L} and a long and detailed discussion with Francesca Matteucci (private communication). We apply an almost identical set of assumptions and observational constraints. \cite{2019arXiv191108450L} is using a code by \citet{2004MNRAS.351.1338L} with their implementation of the IGIMF-R14 and in this work we use the publicly available code GalIMF develop in our previous work \citep{2019A&A...629A..93Y} following a very similar set of assumptions and input parameters as \citet{2019arXiv191108450L}. While the two codes are not identical, they should yield comparable results. However, it appears we draw contradictory conclusions.
Here we discuss the differences in the model, the fitting routine, the results, and the interpretation.

\subsection{DTDs}\label{sec: DTD}

As mentioned in Section~\ref{sec: The DTD of SNIa events}, we test the DTD formulation of the single-degenerate SNIa model formulated in \citet[their eq. 2 to 5]{2001ApJ...558..351M} to demonstrate the effect of applying an alternative DTD.

A different DTD model indeed affects the galactic mean metallicity significantly. However in a way that the observational value is in between the IGIMF-R14 and IGIMF-R14-SD results such that both DTD models are consistent with observation within the $2 \sigma$ uncertainty range.

These tests demonstrate the limitation of the chemical evolution models, where the conclusion depends on the yet unknown nature of the SNIa. Although, in the case of testing the IGIMF-R14 formulation, it happens not to change our conclusion that the IGIMF-R14 model reproduces the Bo\"otes~I data well (and naturally).

\subsection{Set the wind efficiency}\label{sec: Wind efficiency}

The wind efficiency parameter, $\omega$, defined in Section~\ref{sec: Galaxy chemical evolution model}, affects the stellar mass formed after the onset of the galactic wind. A lower wind efficiency leads to a higher galaxy stellar mass and metallicity. 

In \citet{2019arXiv191108450L}, there is written that the wind rate is proportional to the SFR but in reality it is proportional to the amount of gas (Matteucci private communication). This was applied in \citet{2011A&A...531A.136Y} to prevent the wind from stopping when the SFR goes to zero. Thus, we cannot adopt directly the claimed $\omega$ value from \citet{2019arXiv191108450L} but have to find the equivalent $\omega$ that can reproduce their demonstrated SFH.
We tested different values of $\omega$ and find that $\omega \approx 100$ results in a similar SFH.

That means, the $\omega$ is applied only to have a fair comparison with \citet{2019arXiv191108450L}. The strong loss of gas and the fast quenching of the SFH is justified by the stellar age and abundance distribution of the UFDs, as is demonstrated in, e.g., \citet{2014MNRAS.441.2815V}, but the physical mechanism for losing the gas is not necessarily a galactic wind even though we name $\omega$ the "galactic wind efficiency parameter". Thus, the appropriate value of $\omega$ as well as whether the "galactic wind" should depend on the SFR or the remaining gas mass is unsettled and unconstrained.

It has been suggested by \citet{2015MNRAS.446.4220R,2019A&A...630A.140R} that stellar feedback is not effective in removing all the gas from Bo\"otes~I and an external mechanism, such as tidal or ram-pressure stripping, is needed. With the assumed formulation that the galactic wind rate is proportional to the SFR, any mechanism that leads to the gas depletion of Bo\"otes~I is parameterised by a high $\omega$ value.

The proper value of $\omega$ is unknown and should depend on the gwIMF and should therefore be time-dependent. The best-fit solution can be affected when $\omega$ takes a different value or if the gas is removed from the galaxy not by the galaxy wind but by an alternative mechanism \citep{2015MNRAS.446.4220R,2019A&A...630A.140R} or if the suppression of star formation is not due to the gas depletion \citep{2016Natur.535..523F,2017A&A...602A..45L}. 

\subsection{The fitting routine}

As a first step, we made sure that when we apply the same underlying assumptions and input parameters our results and the results of \citet{2019arXiv191108450L} are mutually consistent. 
The low-metal part of the [$\alpha$/Fe]--[Fe/H] relations and especially the Ca abundance are not identical and require further investigation but these differences mainly demonstrate the limitations/uncertainties of chemical evolution models and would not affect our conclusions. This is because it is not a challenge for a model to fit the [$\alpha$/Fe]-[Fe/H] relations, as this can almost always be done by tuning the gas-depletion timescale.
The main challenge, on the other hand, is to simultaneously reproduce the observed $M_{\rm *,final}$ and $\mathrm{[Fe/H]}_{\rm mean}$ (or MDF, in the case of \citealt{2019arXiv191108450L}, see Section~\ref{sec: The number of SNIa events} and \ref{sec: Mean stellar [Fe/H]}) of the galaxy.


As it is explained in Section~\ref{sec: Input parameters}, there are two free input parameters while there are three independent observables. We fit $M_{\rm *,final}$ within $1\sigma$ deviation of the observation by tuning the input parameter $M_{\rm ini}$ while the $\mathrm{[Fe/H]}_{\rm mean}$ is left to test the model.
It turns out that the resulting $\mathrm{[Fe/H]}_{\rm mean}$ of our model applying the IGIMF theory naturally agrees with the observational values within about $2 \sigma$, representing the observations better than the models shown in \citet{2019arXiv191108450L}.

The fitting routine explained in Section~\ref{sec: Input parameters} allows us to identify those input parameters that result in a good agreement with the data. While \citet{2019arXiv191108450L} studies a sparse grid of input parameter sets, there do not contain or cover the best fitting input parameters we found. This seems to be the main reason that we end up with different conclusions.


\subsection{The [$\alpha$/Fe]-[Fe/H] relations}\label{sec: The [alpha/Fe]-[Fe/H] relation}

As is shown in Fig~\ref{fig:MgSiCa_3}, the intrinsic scatter of the [$\alpha$/Fe]-[Fe/H] data is large due to the complex galactic gas and metal distribution and stars may form outside the main progenitor halo \citep{2017ApJ...848...85J}. Our mean galactic metal evolution track assumes that all the gas is always well mixed, and thus cannot explain the metal abundance of individual single stars.

The best-fit model applying the IGIMF theory is comparable with the solutions given by \citet{2019arXiv191108450L} shown as the dashed and dash-dotted lines in Fig~\ref{fig:MgSiCa_3}, especially for metal-rich stars. But the highest [$\alpha$/Fe] value at the metal-poor end of the plots from \citet{2019arXiv191108450L} is smaller than our results and fit the data not as well as our model. 

\begin{figure}
    \centering
    \includegraphics[width=\hsize]{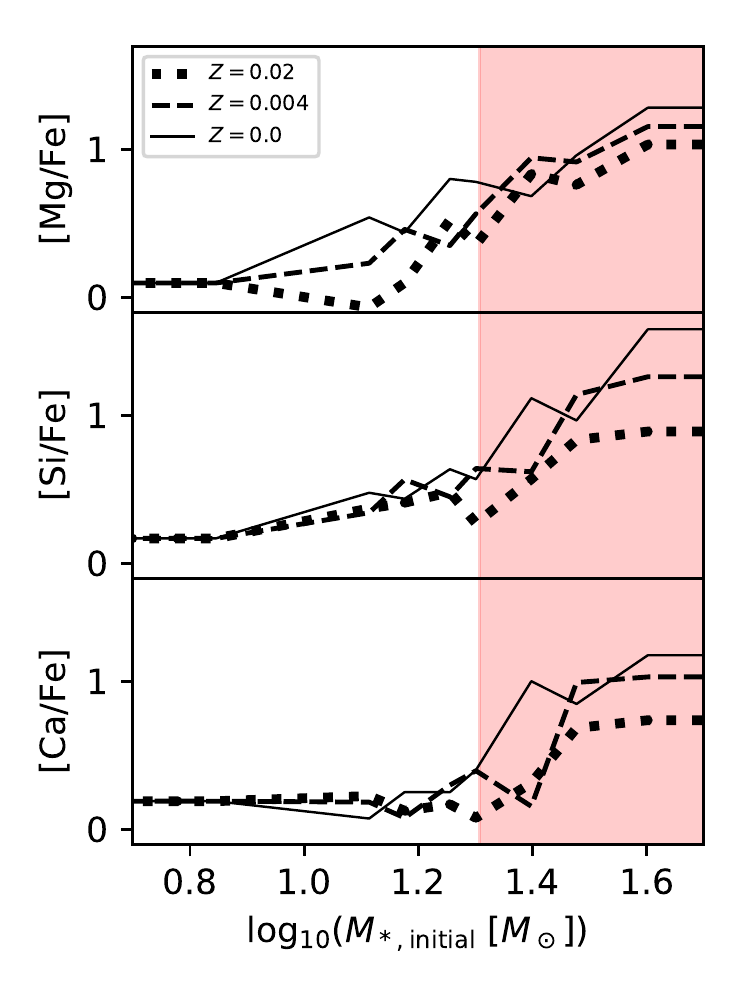}
    \caption{Stellar metal yield ratios of stars with different mass and metallicity given by \citet{2006ApJ...653.1145K}. The yields from \citet{2006ApJ...653.1145K} for massive stars are connected to the yields for low- and intermediate-mass stars as given by \citet{2001A&A...370..194M}. The yields for $M_*>40\,M_\odot$ are the same as the yields for $M_*=40\,M_\odot$ stars. 
    The red shaded region with log$_{10}(M_{\mathrm{*, initial}}[M_\odot])>1.3073$ indicates the zero metallicity stars with a lifetime shorter than 10 Myr according to \citet{2019A&A...629A..93Y}.
    }
    \label{fig:stellar_yields}
\end{figure}

The highest [$\alpha$/Fe] value should be the IMF-weighted [$\alpha$/Fe] value of all the massive stars with Z=0 and a lifetime shorter than the first timestep, $\delta t=10$ Myr, that is, the IMF weighted thin solid line within the red shaded region of Fig.~\ref{fig:stellar_yields}. Since the IMF weighted value should be higher than the lowest value in the red region, it is clear that the initial [Mg/Fe] should be higher than 0.5, which agrees with our result and disagrees with \citet{2019arXiv191108450L}.

The [$\alpha$/Fe] plateau can be enhanced by a more complicated gas-flow model, a larger stellar-mass bin, or a larger model timestep. A non-simultaneous star formation combined with local stellar wind pollution can also complicate the model prediction, which requires more elaborate hydrochemical simulations. 

We note that the metal-poor part of the [$\alpha$/Fe]--[Fe/H] relations is affected by the small number of stars formed at the earliest timesteps. Therefore, the metal-poor part is not reliable anyway due to the sheer small number of massive stars affecting it, the uncertainty of the stellar winds of these massive stars, and whether or not they explode as supernovae. The stars more massive than $40 M_\odot$ share the same adopted stellar yield table, which is certainly not the ideal setup to discuss the chemical abundance of the extremely metal-poor era of a galaxy.

\subsection{Star formation history}\label{sec: Star formation history}

The tested SFT in \citet{2019arXiv191108450L} is generally longer than our best-fit models.

Since the gas flow and star formation criteria in the chemical evolution model are uncertain, the best-fit SFH is not conclusive. Here we compare it with the independently estimated SFH given by the CMD. \citet{2014ApJ...796...91B} demonstrate that the CMD of Bo\"otes~I is best-fitted by, essentially, a single starburst in a short timescale, in agreement with our result. We note that the cumulative SFH shown in \citet[their fig. 8]{2014ApJ...796...91B} is consistent with their best-fit model and our resulting SFH as well.

Thus, the short SFT required by our model to fit the [$\alpha$/Fe]-[Fe/H] relations is supported by the SFH obtained from the CMD of Bo\"otes~I \citep{2014ApJ...796...91B} and is consistent with the average gas-depletion timescale of dwarf galaxies \citep{2009ApJ...706..516P}. Our results are also in line with \citet{2015ApJ...799L..21W} showing that the shortest SFT required for the observed self-enrichment of the UFDs is about 0.1 Gyr.
The gas of low-mass-satellite galaxies was removed by their interaction with the Milky May \citep{2015MNRAS.446.4220R,2019A&A...630A.140R} such that they shut-off star formation on a short timescale, as also found for Dragonfly 44 by \citet{2019ApJ...884L..25H}.

\begin{figure}
    \centering
    \includegraphics[width=\hsize]{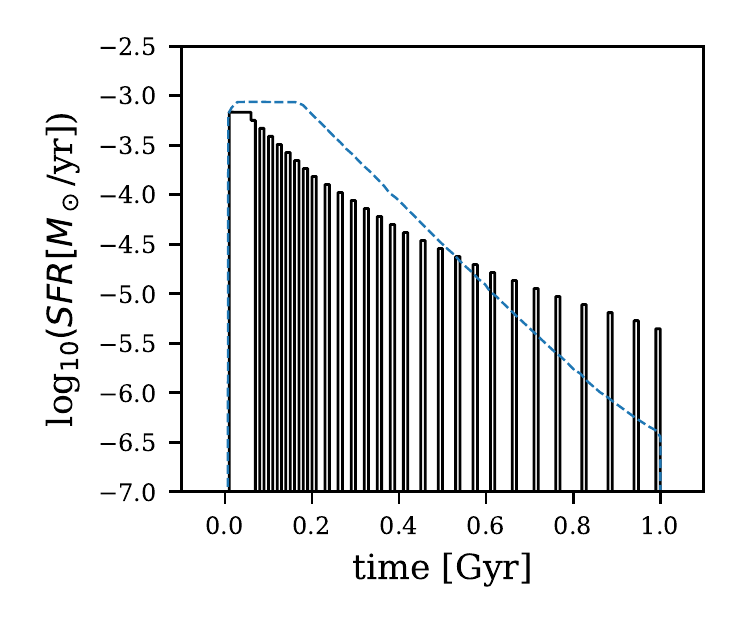}
    \caption{A SFH requiring an artificial gas-cooling time. The blue-dashed line is the same as in Fig.~\ref{fig:SFH_3}. Compared to the best-fit IGIMF-R14 model (with a SFH shown in Fig.~\ref{fig:SFH_3}), the here discretized SFH has a larger spread but similar galaxy evolution results (see Section~\ref{sec: Star formation history}).}
    \label{fig:SFH_4}
\end{figure}

In addition, we note that when the SFR drops low, the continued star formation activity does not have much of an effect on the chemical abundances. If we assume that, after the onset of the galactic wind and after each star formation epoch, the extremely diluted gas needs one to a few 10 Myr cooling time before forming new stars \citep{2017A&A...602A..45L}, then the SFH may be greatly extended, such as the one shown in Fig.~\ref{fig:SFH_4}. But the resulting [$\alpha$/Fe]-[Fe/H] relation, $M_{\rm *,final}$, and $\mathrm{[Fe/H]}_{\rm mean}$ are barely affected.

\subsection{Interpretation of the result}\label{sec: Interpretation of the result}

Since there is still a $2 \sigma$ mismatch between the model prediction and the observation, the question is whether the $2 \sigma$ difference excludes the assumed IGIMF-R14 formulation.
We consider it an acceptable agreement because the uncertainty of the assumptions applied in the model (see below) can certainly cause a $2 \sigma$ disagreement, i.e., the discrepancy may not be due to a wrong gwIMF assumption but errors in other assumptions.

For example, a lower dark matter mass and/or a higher $\eta_{\rm SNII}$ (0.1 instead of the 0.03 applied here, see Section~\ref{sec: Galaxy chemical evolution model}, is possible, \citealt{1998A&A...337..338B,2015MNRAS.446.4220R}) combined with a higher $M_{\rm ini}$ would help reduce the value of [Fe/H]$_{\rm mean}$. Metal-rich gas ejected by the supernovae may be preferentially expelled from such a low-mass galaxy especially for off-centre explosions \citep{2014ApJ...796...11W,2015MNRAS.446.4220R,2019A&A...630A.140R}. Similarly, the galaxy's radius at the onset time of the galactic wind may be different from its present-day radius. The DTD of SNIa explosions is uncertain and may peak at a different time (see Section~\ref{sec: DTD}). The stellar yield may be different and the massive stars may collapse to black holes without ejecting metals to the gas. Other assumptions in the model, such as the single-zone assumption, instantaneous-gas-mixing assumption, etc., all affect the fitting results and it is difficult to quantify how large the effect is.

\subsection{Summary}

In summary, we first reproduce the results of \citet[their model 3BooI]{2019arXiv191108450L} with the same assumptions and input parameters applied to make sure the results from the two code are in line and comparable. Then we find a set of input parameters that lead to a better model-fit with the observations than the models in \citet{2019arXiv191108450L}. Finally, we discuss the various observational and model uncertainties and conclude that, with the current accuracy, the IGIMF model is describing the observed properties of Bo\"otes~I self-consistently and naturally, and in particularly better than the solution assuming an invariant Salpeter IMF shown in \citet{2019arXiv191108450L}.

\section{Conclusion}\label{sec: Conclusion}

By adopting independently measured galaxy parameters for Bo\"otes~I we are able to limit the number of free parameters in the galaxy chemical evolution models and test different IMF formulations. 

By varying only two free parameters, the gas-depletion timescale and the initial gas mass that together determine the SFH, all the observed properties of Bo\"otes~I can be well reproduced assuming the IGIMF theory originally formulated by \citet{2003ApJ...598.1076K}, in contrast to the conclusion stated by \citet{2019arXiv191108450L}. Comparing with the best-fit solution assuming the Salpeter IMF in \citet{2014MNRAS.441.2815V} and \citet{2019arXiv191108450L}, the IGIMF model applying the IGIMF theory agrees better with the data. 

More explicitly, as listed in Table~\ref{tab:parameter}, the IGIMF-R14 formulation assuming both power-law and single-degenerate DTDs (i.e., our IGIMF-R14 and IGIMF-R14-SD models) are consistent with the data at the $2\sigma$ confidence level. Formulation IGIMF(A2) and IGIMF(A3) are in disagreement with the data at the $4\sigma$ confidence level, while the IGIMF(A4) formulation fits very well.

We emphasise that when applying the IGIMF theory to a galaxy, such as Bo\"otes~I here, the parameters are not varied at liberty to force a good fit. Rather, the parameters (e.g. the IMF shape, the gas consumption timescale, the mass, the DTD) are significantly constrained by independent sources or data. Therefore, it is most remarkable that Bo\"otes~I, which represents a rather extreme star-forming system, comes out to be so well-described using the IGIMF
theory. Such an agreement is, of course, expected if the applied theory is relevant for describing the observed physical phenomena. 

The chemical abundance of the dwarf galaxy combined with the IGIMF theory (i.e. a computed top-light gwIMF) suggests a short SFT of about 0.1 Gyr (similar to the dwarf galaxy Dragonfly 44, \citealt{2019ApJ...884L..25H}). This not only agrees well with the stellar synthesis study of Bo\"otes~I \citep{2014ApJ...796...91B} but also is self-consistent with the SFR--galaxy-gas-mass measurements (i.e. the gas-depletion timescale of galaxies) when the IGIMF theory is assumed \citep{2009ApJ...706..516P}. We emphasize that the long gas-depletion timescale suggested by earlier studies solely depends on the assumption that Bo\"otes~I has a Salpeter IMF.

A more devoted study considering detailed features of individual dwarf galaxies (e.g. the stellar metallicity distribution function, other chemical elements, etc.) and taking into account more galaxies is required to further test the stellar population and chemical evolution models. 
The capability of such a chemical evolution study is mainly limited by the uncertainty on how the gas is accreted, mixed, and expelled. The currently applied assumption that the gas is always instantaneously well-mixed may not be appropriate for the dwarf galaxies. 

Due to these and other uncertainties discussed above, simple galaxy chemical evolution models can provide suggestions and insights but still cannot provide conclusive definite constraints on the IMF model (cf. \citealt{2016IAUS..317..164R}).

In summary:
\begin{itemize}
  \item From CMD synthesis we know that the stars in Bo\"otes~I are old and formed in a short timescale. Thus only the IMF for massive stars (with stellar lifetimes < SFT) determines the [$\alpha$/Fe]--[Fe/H] relation. The top-light gwIMF predicted by the IGIMF theory accounts better for the [$\alpha$/Fe]-[Fe/H] data than the Salpeter gwIMF. The solution agrees with the gas-depletion timescale for star-forming galaxies assuming the IGIMF theory \citep{2009ApJ...706..516P} and evidence from H$\alpha$/UV luminosity ratios for dwarf galaxies \citep{2009ApJ...706..599L} naturally. The observed deficit of massive stars in resolved dwarf galaxies is accounted for very well by the IGIMF theory \citep{2018MNRAS.477.5554W}. Therefore, this evidence together robustly confirms the gwIMF being top-light in dwarf galaxies.
  \item With the top-part of the gwIMF and its time variation determined, we have three variables (the gas-depletion timescale, the initial gas mass, the bottom-IMF slope) and three constraints (the observed [$\alpha$/Fe]--[Fe/H] data, the observed galaxy mass, the observed mean stellar metallicity). The gas-depletion timescale is almost solely determined by the [$\alpha$/Fe]--[Fe/H] relations (and the top-IMF shape); the initial gas mass is then determined by the observed galaxy mass and gas-depletion timescale. Therefore, the bottom part of the gwIMF can be, in principle, constrained by the observed mean metallicity. We find that the IGIMF-R14 formulation naturally fits the observation with the two different SNIa DTDs we tested (i.e. the IGIMF-R14 and IGIMF-R14-SD models). However, if the gwIMF is more top-light as suggested by the IGIMF(A2) and IGIMF(A3) formulations, a mildly\footnote{The preferred IMF formulation, i.e., IGIMF(A4), is in between the IGIMF(A2) formulation with a canonical bottom-IMF and the IGIMF(A3) formulation with a bottom-light IMF.} bottom-light IMF for sub-solar metallicity is preferred. We propose a bottom-IMF formulation (Eq.~\ref{eq: alpha new}) which accommodates these findings in Section~\ref{sec: Other IGIMF formulations}, i.e. model IGIMF(A4).
  \item Our IGIMF-R14-SD model is largely consistent with the code applied by \citet{2019arXiv191108450L} when the same input parameters are applied while the remaining disagreement of the predictions from the two codes (e.g. the metal-poor part of the [$\alpha$/Fe]-[Fe/H] relations, Fig.~\ref{fig:MgSiCa_5}) mainly demonstrates the limitation and uncertainty of chemical evolution models. Our model fits the observed $M_{\rm *,final}$ and $\mathrm{[Fe/H]}_{\rm mean}$ within $2\sigma$ and we interpret this result as an acceptable agreement considering the large uncertainty of the assumptions applied in the model.
\end{itemize}

In general, the UFDs formed stars at extreme physical conditions and thus provide promising pathways to future tests and investigations of chemical evolution and the environment-dependent IMF. 

The evolution modelling of Bo\"otes~I based on a systematically evolving gwIMF demonstrates this dwarf galaxy to not be unusual. It follows the same gas-consumption process as other galaxies but lost its gas supply within about 0.1 Gyr due to its high-energy orbit around the Milky Way.

\begin{acknowledgements}
We thank Francesca Matteucci for a long and detailed discussion on the differences between this work and that of \citet{2019arXiv191108450L} from which this work has benefitted. We also thank Donatella Romano for helpful discussions. ZY acknowledges financial support from the China Scholarship Council (CSC, file number 201708080069). TJ acknowledges support by the Erasmus+ programme of the European Union under grant number 2017-1-CZ01- KA203-035562. The development of our chemical evolution model applied in this work benefited from the International Space Science Institute (ISSI/ISSI-BJ) in Bern and Beijing, thanks to the funding of the team “Chemical abundances in the ISM: the litmus test of stellar IMF variations in galaxies across cosmic time” (Donatella Romano and Zhi-Yu Zhang).
\end{acknowledgements}

\bibliographystyle{aa} 
\bibliography{references}

\begin{appendix}

\section{Results with a different DTD}\label{sec: other_results}

The results of model IGIMF-R14-SD, assuming the single-degenerate DTD formulation, are documented here. See Section~\ref{sec: The DTD of SNIa events} for a description of the model.
Other input and output parameters of the model are listed in Table~\ref{tab:parameter}. 

\begin{figure}
    \centering
    \includegraphics[width=\hsize]{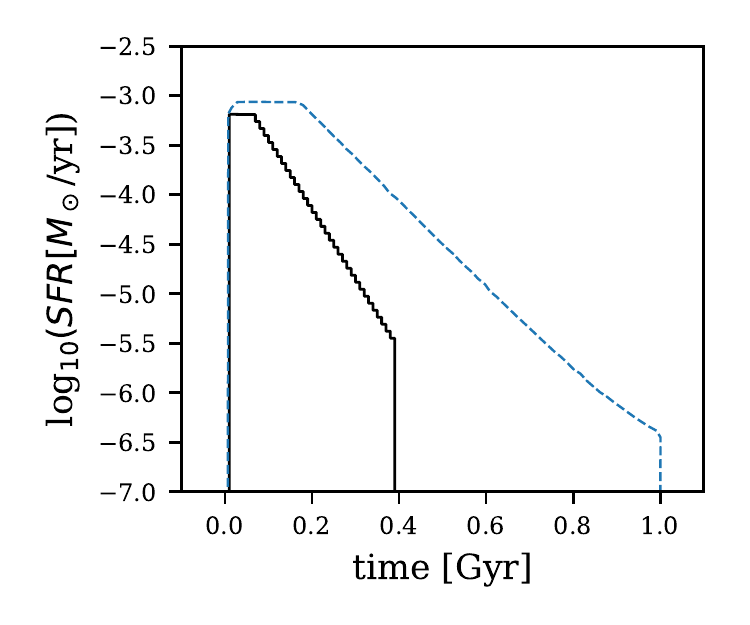}
    \caption{Same as Fig.~\ref{fig:SFH_3} but for the IGIMF-R14-SD model.}
    \label{fig:SFH_5}
\end{figure}

\begin{figure}
    \centering
    \includegraphics[width=\hsize]{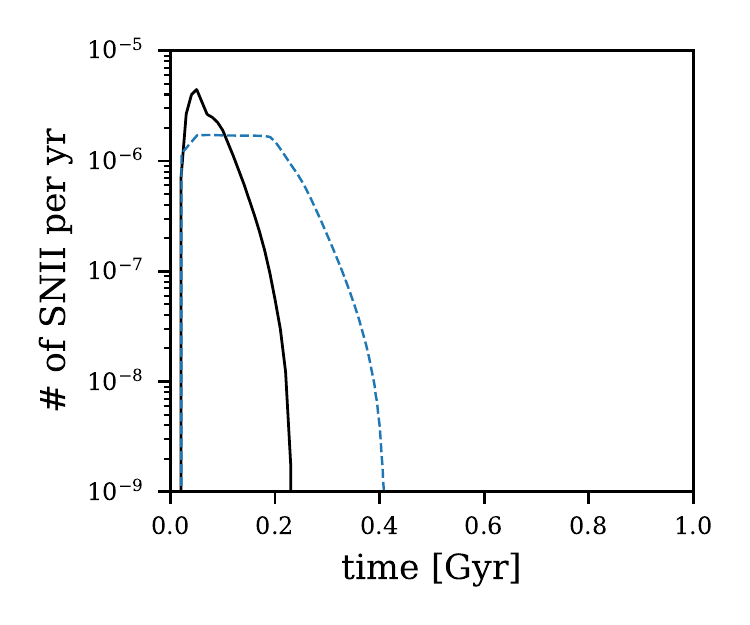}
    \caption{Same as Fig.~\ref{fig:SNII_3} but for the IGIMF-R14-SD model.}
    \label{fig:SNII_5}
\end{figure}

\begin{figure}
    \centering
    \includegraphics[width=\hsize]{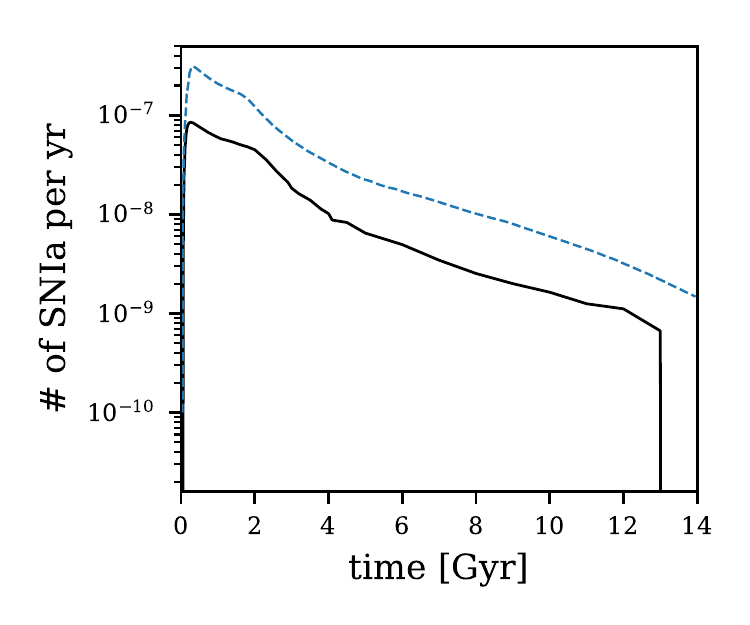}
    \caption{Same as Fig.~\ref{fig:SNIa_3} but for the IGIMF-R14-SD model. The SNIa events stop at about 13 Gyr due to the assumption that $M_2>0.8 M_\odot$ following \citet{2001ApJ...558..351M}.}
    \label{fig:SNIa_5}
\end{figure}

\begin{figure}
    \centering
    \includegraphics[width=\hsize]{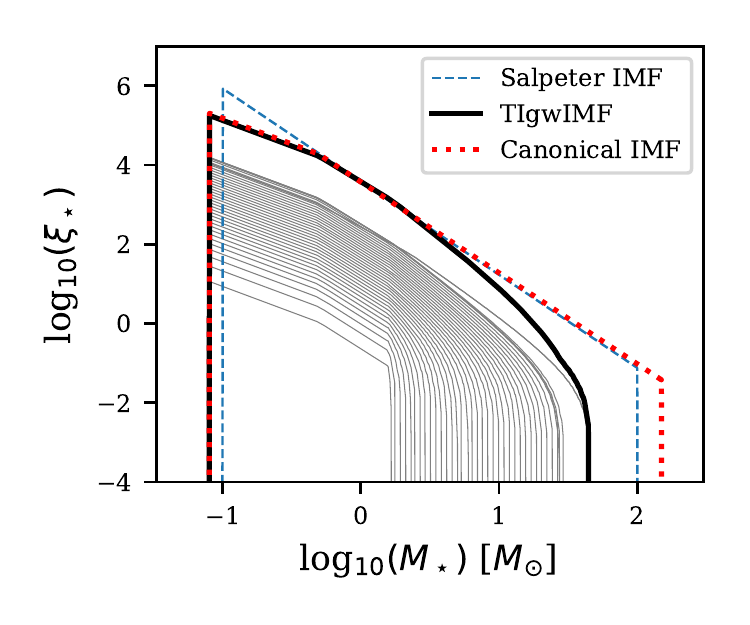}
    \caption{Same as Fig.~\ref{fig:TIgwIMF_3} but for the IGIMF-R14-SD model.}
    \label{fig:TIgwIMF_5}
\end{figure}

\begin{figure}
    \centering
    \includegraphics[width=\hsize]{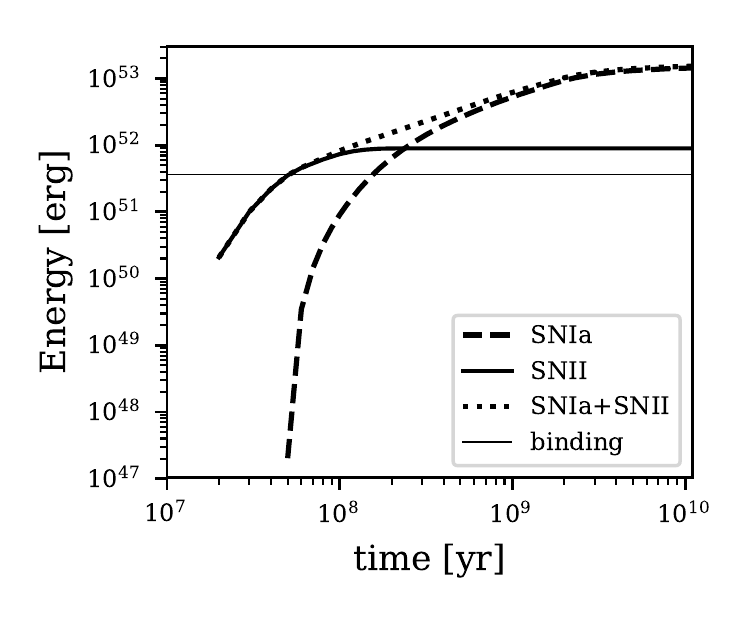}
    \caption{Same as Fig.~\ref{fig:energy_evolution_3} but for the IGIMF-R14-SD model.}
    \label{fig:energy_evolution_5}
\end{figure}

\begin{figure}
    \centering
    \includegraphics[width=\hsize]{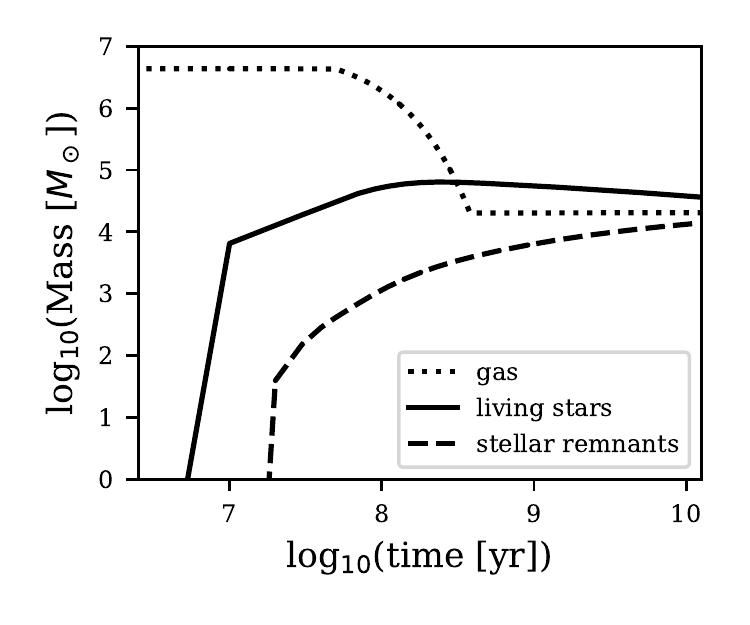}
    \caption{Same as Fig.~\ref{fig:mass_evolution_3} but for the IGIMF-R14-SD model.}
    \label{fig:mass_evolution_5}
\end{figure}

\begin{figure}
    \centering
    \includegraphics[width=\hsize]{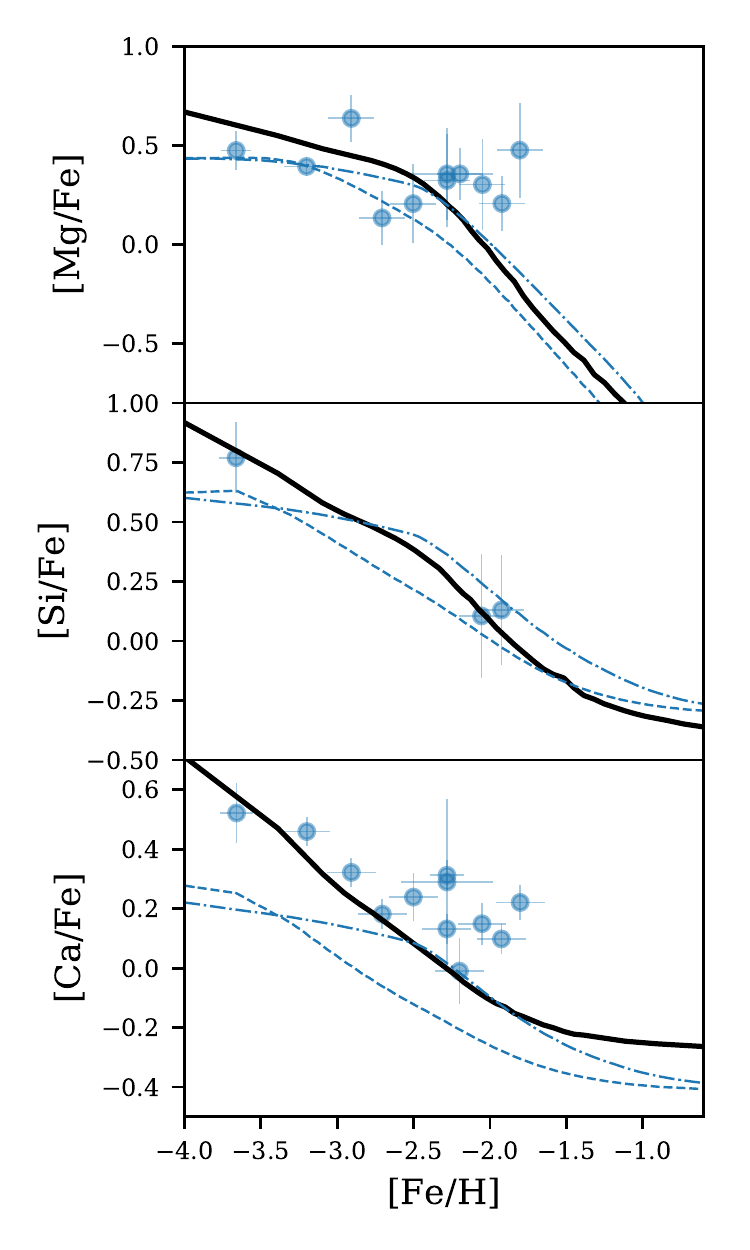}
    \caption{Same as Fig.~\ref{fig:MgSiCa_3} but for the IGIMF-R14-SD model.}
    \label{fig:MgSiCa_5}
\end{figure}

\end{appendix}

\end{document}